\documentclass{article}
\input{header.sty}
\usepackage{arxiv}

\usepackage[utf8]{inputenc} 
\usepackage[T1]{fontenc}    
\usepackage{hyperref}       
\usepackage{url}            
\usepackage{booktabs}       
\usepackage{amsfonts}       
\usepackage{nicefrac}       
\usepackage{microtype}      
\usepackage{cleveref}       
\usepackage{lipsum}         
\usepackage{graphicx}
\usepackage{doi}
\usepackage{natbib}
\usepackage{amsthm}

\title{A class of modular and flexible covariate-based covariance functions for nonstationary spatial modeling}

\date{}

\newif\ifuniqueAffiliation
\uniqueAffiliationtrue

\ifuniqueAffiliation 
\author{ \href{https://orcid.org/0000-0001-9337-7154}{\includegraphics[scale=0.06]{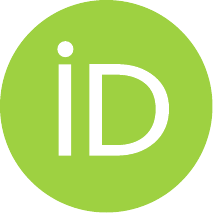}\hspace{1mm}Federico Blasi} \\
	Department of Mathematical \\ Modeling and Machine Learning\\
	University of Zurich\\
	Zurich, Switzerland \\
	\texttt{federico.blasi@uzh.ch} \\
	\And
	\href{https://orcid.org/0000-0002-6319-2332}{\includegraphics[scale=0.06]{orcid.pdf}\hspace{1mm}Reinhard Furrer} \\
	Department of Mathematical \\ Modeling and Machine Learning \\
University of Zurich\\
	Zurich, Switzerland \\
	\texttt{reinhard.furrer@math.uzh.ch} \\
}
\else
\usepackage{authblk}

\newbox{\orcid}\sbox{\orcid}{\includegraphics[scale=0.06]{orcid.pdf}} 
\fi

\renewcommand{\shorttitle}{A Class of Modular and Flexible Covariate-based Covariance Functions}

\hypersetup{
pdftitle={A class of modular and flexible covariate-based covariance functions for nonstationary spatial modeling},
pdfauthor={Federico Blasi},
}

\newtheorem{lemma}{Lemma}

\begin{document}
\maketitle

\begin{abstract}
	 Paradoxically, while the assumptions of second-order stationarity and isotropy appear outdated in light of modern spatial data, they remain remarkably robust in practice, as nonstationary methods often provide marginal improvements in predictive performance.
    This limitation reflects a fundamental trade-off: nonparametric approaches, while offering extreme flexibility, require substantial tuning to avoid overfitting and numerical challenges in practice, while parametric approaches are more robust against overfitting but are constrained in flexibility, often facing considerable numerical challenges as flexibility increases. 
    In this article we introduce a parametric class of covariance functions that extends the use of parametric nonstationary spatial models, aiming to compete with the flexibility and local adaptability of nonparametric approaches. The covariance function is modular in the sense that allows for separate parametric structures for different sources of nonstationarity, such as marginal standard deviation, geometric anisotropy, and smoothness. The proposed covariance function retains the practical identifiability and computational stability of parametric forms while closing the performance gap with fully nonparametric methods. 
    A Mat\'ern stationary isotropic model is nested within the complex model and can be adapted such that it is computationally feasible for handling thousands of observations. 
    A two-stage approach can be employed for model selection.
    We explore the statistical properties of the presented approach, demonstrate its compatibility with the frequentist paradigm, and highlight the interpretability of its parameters.
    We illustrate its prediction capabilities as well as interpretability through an analysis of Swiss monthly precipitation data, showing that Gaussian process models with the presented covariance function, while remaining robust against overfitting, provide quantitative and qualitative improvements over existing approaches.
\end{abstract}


\section{Introduction}\label{sec1}
Gaussian process models provide a fundamental framework for geostatistical analysis of spatial data. Its key component, the covariance function, has traditionally been assumed to be stationary, implying consistent covariance across spatial distances, regardless of location. However, in light of modern spatial data, the assumption of stationarity has become increasingly difficult to justify, calling for more flexible approaches. 
Various methods have been developed to overcome the rigidity of stationary covariance functions, including convolving stationary orthogonal processes \citep{fuentes2001high}, applying deformation techniques \citep{sampson1992nonparametric}, and leveraging deep learning methods \citep{zammit2021deep}. For comprehensive overviews of nonstationary approaches, see \cite{gelfand2010handbook} (Ch. 9), \cite{fouedjio2017second}, and \cite{schmidt2020flexible}. 

In this article, we focus on a class of nonstationary covariance functions that explicitly incorporate spatial information, capturing deviations from stationarity based on spatial characteristics. Similar to mean regression, by conditionally modeling the covariance on observed covariates, we can efficiently capture the spatially-varying nature of the spatial structure through economical parameterizations (i.e., parsimonious). The explicit relationship between covariates and deviations from stationarity enhances interpretability, offering insights into how spatial characteristics shape the underlying process.

Covariate-based covariance functions have been actively researched during the last two decades.
\cite{hoef2006spatial} developed spatial models whose covariance structures incorporate flow and stream distance through spatial moving averages. 
\cite{cooley2007bayesian} modeled extreme precipitation by representing the process in the climate space, mainly composed by elevation and mean precipitation at the weather station.
\cite{calder2008dynamic} included wind direction information from a single location in the kernel convolution approach of \cite{higdon1998process}. 
\cite{reich2011class} extended the work done by \cite{fuentes2002spectral}, modeling the nonstationarity covariance function as a weighted sum of independent stationary zero-mean Gaussian processes, where the weights are obtained through spatially related covariate information.
\cite{schmidt2011considering} extended the work done by \cite{sampson1992nonparametric} considering a $d$-dimensional space, from which $d-2$ are related to covariates. The stationary isotropic covariance function of the extended space is modeled with a Mat\'ern covariance function with a Mahalanobis distance that models the roughness and smoothness of the spatial process for the different directions. 
Another extension of \cite{sampson1992nonparametric} is the work done by \cite{bornn2012modeling}, who devised a method that embeds the original nonstationary field in a higher-dimensional space where it can be more straightforwardly described and modeled. It differs from the work done by \cite{sampson1992nonparametric} in that here, the locations in the geographic space are retained, with added flexibility obtained through the extra dimensions related to covariates.
\cite{ingebrigtsen2014spatial} represented nonstationarity in the second-order through covariates, as proposed by \cite{lindgren2011explicit} where it is shown that a Gaussian random field with a Mat\'ern covariance function can be represented as the stationary solution of a linear stochastic partial differential  equation (SPDE). 
\cite{neto2014accounting} modeled the covariance structure conditionally on the wind direction information for an air pollution process. This is done via the convolution approach, proposing tailored functions that include wind direction.
\cite{risser2015regression} introduced a covariance function based on the nonstationary covariance model of \cite{paciorek2006spatial} that considers covariate information. Sources of nonstationarity such as the marginal standard deviation and spatial anisotropy are modeled separately with a parametric model. The focus is mainly on interpretability while preserving a low-dimensional parameterization, where the flexibility component is sensitive to the covariates and the parametric model offered.
\cite{gilani2016non} presented a nonstationary spatio-temporal model for three traffic-related pollutants in a localized near-road environment, combining the nonstationary methods by \cite{fuentes2002spectral} and \cite{schmidt2011considering}, each of them considering covariates such as distance from the main road and wind direction, among others, driving the nonstationarity and the mixture weights.
\cite{xu2018improved} proposed an improved latent space approach for modeling nonstationary spatial and spatiotemporal random fields. By considering regressors as latent dimensions, they characterize the nonstationarity using a regressor-based standard deviation and correlation.

While all these methods can accommodate nonstationarity, each comes with certain limitations. Many are tailored to specific data contexts or phenomena (e.g. the stream network model of \cite{hoef2006spatial} or the traffic pollution model of \cite{gilani2016non}), limiting their broader applicability, or present computational challenges \citep{neto2014accounting}, \citep{risser2015regression}, \citep{ingebrigtsen2014spatial}. The added flexibility often comes at the cost of heavy computations or numerical instability, as noted for the wind-informed convolution model of \cite{neto2014accounting}, the covariate-dependent model of \cite{risser2015regression}, and the SPDE-based approach of \cite{ingebrigtsen2014spatial}.
A common theme is that achieving greater flexibility in the covariance function typically incurs substantially higher computational and implementation complexity. This steep trade-off, combined with significant technical overhead, has often discouraged practitioners from adopting nonstationary covariance models in practice. As a result, nonstationary covariance approaches remain less popular than expected, despite their potential to improve predictions.
However, more flexibility does not always translate into better prediction. If the data cannot support many local parameters, because of limited sample size or high measurement noise, overly flexible models can increase out‐of‐sample error. Thus, spatial models should strike a balance between expressiveness and parsimony to achieve robust, stable predictions in practice.

Our article presents a class of covariate-based covariance functions that provides a convenient tradeoff, by offering a flexible and economical representation of the nonstationary process, representing a wide range of key sources of nonstationarity of the spatial structure in a modular framework. 
It allows for separate parametric structures for different types of nonstationarity, such as variance, local anisotropy, as well as smoothness. 
This modular parametric structure can be leveraged to perform efficient model selection alongside parameter estimation, helping identify which covariates (and which aspects of the covariance) contribute meaningfully to the fit.
It simplifies to a Mat\'ern covariance function in its basic form and, thanks to its modularity, is adaptable for large datasets, extending the convenient tradeoff across a wide range of sample sizes. 
We investigate the proposed covariance function in the challenging setting of a single realization of a spatial process observed over a bounded domain. In this context, we discuss interpretability and examine the potential pitfalls and benefits of using such flexible covariance functions as spatial smoothers.

The article is structured as follows. In Section~\ref{ch:sec2}, we introduce the likelihood-based framework as well as the stem of nonstationary covariance functions based on convolution, necessary roots to introduce Section~\ref{ch:consider_covs}, where we present the modular nonstationary covariance function and explore its interpretability and discuss potential challenges of covariate-based nonstationary covariance functions. 
In Section~\ref{ch:illu}, we apply the proposed approach to Swiss monthly precipitation data, which exhibit highly heterogeneous spatial structures, driven by complex orography and climatic gradients, that are well known to induce nonstationarity in environmental fields (e.g. \citealp{paciorek2006spatial}; \citealp{ingebrigtsen2014spatial}; \citealp{risser2015regression}). 
Finally, Section~\ref{ch:sec5} concludes with a summary of our findings and directions for future work.

\section{Likelihood-based approach with nonstationary covariance functions} \label{ch:sec2}
It is of common practice to assume that the spatial variable $Z$ defined on the study region $\cD \subseteq \mathcal{R}^d$, can be modeled as a Gaussian process $Z(\cdot)\sim \mathrm{G}\mathrm{P}(\mu(\cdot),\Cov(\cdot,\cdot))$ with some mean function $\mu(\cdot)$ and covariance function $\Cov(\cdot,\cdot)$. Considering $\bs\in\cD$ as the spatial location, a typical decomposition of $Z(\cdot)$ is then given by
\begin{equation}
     Z(\bs) = \mu(\bs) + Y(\bs) + \epsilon(\bs) \ , \ \  \bs \in \cD,
\end{equation}
where $Y(\cdot)$ is a zero-mean continuous Gaussian process representing the spatial dependencies, and where $\epsilon(\cdot)$ is often considered to describe the measurement error and small-scale variability, represented as a Gaussian random noise process with mean zero and variance $\sigma^2_{\epsilon}$, independent of $Y(\cdot)$.
We assume that our sample $\bz =(z_1,\dots,z_n)^{\T}$ is the result of observing $Z(\cdot)$ at mutually distinct sampling locations $\{\bs_1,\dots,\bs_n\}$, i.e., observing a multivariate Gaussian distribution $\{Z(\bs_1),\dots, Z(\bs_n)\} \sim \mathcal{N}_{n}(\boldsymbol{\mu},\bSigma_{\mathbf{Y}} + \mathbf{I}_n \sigma^2_{\epsilon})$, being $\boldsymbol{\mu}$ a $n\times1$ vector of elements $\mu(\bs_{\ell})$, $\bSigma_{\mathbf{Y}}$ a $n\times n$ symmetric positive semi-definite matrix with elements $\left[\bSigma_{\mathbf{Y}}\right]_{i,j}=\Cov(\bs_i,\bs_j)$, and where $\mathbf{I}_n \sigma^2_{\epsilon}$ is the component associated with the error process $\epsilon(\cdot)$.
Then, $\bSigma_{\Ze} = \bSigma_{\Ye} + \mathbf{I}_n \sigma^2_{\epsilon}$.
For the remainder of the article, we adopt parametric forms for $\mu(\cdot)$ and $\Cov(\cdot,\cdot)$, with $\mu(\cdot;\bbeta)$ and $\Cov(\cdot,\cdot;\bpsi)$, where $\bbeta$ and $\bpsi$ are the associated unknown parameters, vectors of dimension $p+1$ and $m$, respectively. 

The covariance function $\Cov(\cdot,\cdot;\bpsi)$ is often selected from one of the low-dimensional parameterization covariance functions (which we call \textit{classical} covariance functions) that assumes that the underlying Gaussian process $Y(\cdot)$ is stationary, imposing the mean and covariance to be invariant under global shifts in $\mathcal{R}^d$, i.e., $\mu(\bs + \mathbf{h}) = \mu(\bs)$ and that $\Cov(\bs_i + \mathbf{h}, \bs_j + \mathbf{h};\bpsi) = \Cov(\bs_i, \bs_j;\bpsi), \forall \mathbf{h} \in \mathbb{R}^d$. 
Moreover, an often stated and more restrictive assumption is isotropy, imposing that  $\Cov(\cdot,\cdot;\bpsi)$ is a function of $h = ||\mathbf{h}||$ only, where $||\cdot||$ is a norm such as the Euclidean. Then, the process is said to be isotropic.
Among the classical covariance functions, the Mat\'ern family has received a lot of attention over the last two decades \citep{matern2013spatial,porcu2023mat}, and takes the form of
\begin{equation} \label{eq:cov_mat}
    \Cov(h;\bpsi) = \sigma^2\frac{2^{1-\nu}}{\Gamma(\nu)}\Bigl(\sqrt{8\nu}\frac{h}{\gamma}\Bigr)^{\nu}\cK_{\nu}\Bigl(\sqrt{8\nu}\frac{h}{\gamma}\Bigr),
\end{equation}
where $\sigma >0,\gamma >0,\nu >0, \cK_{\nu}(\cdot)$ is the modified Bessel function of the second kind of order $\nu$ \citep{Abramowitz-Stegun:1965}, and $\Gamma(\cdot)$ is the gamma function.
This parameterization links the distance $\gamma$ at which the spatial correlation is approximately $0.1$ \citep{lindgren2011explicit}.
The parameter $\nu$ controls the degree of smoothness of the process, shaping the correlation structure at infinitesimal small distances. 
\cite{banerjee2003smoothness} comment that for essentially featureless areas (i.e., flat surfaces), one would expect continuous and differentiable surfaces, whereas, for areas with irregular features such as ridges or canyons, even continuity would be inappropriate.
As special cases, the Mat\'ern covariance function approaches the Gaussian covariance model when $\nu \to \infty$ (up to a rescaling) and simplifies to the exponential covariance model when $\nu = 1/2$.
\citep{stein1999interpolation} provides in detail the asymptotic convenience of using covariance functions with flexible degree of smoothness.
The flexibility of classical covariance models is often extended by considering a global affine transformation of the Euclidean distance, where instead of the Euclidean distance $h$, we consider the affine transformation ${\Delta\bs}^{\T}A^{-1} {\Delta\bs}$, with $A$ a $2\times 2$ symmetric non-singular matrix and $\Delta\bs = \bs_i - \bs_j$. In this scenario, we say that the process is geometrically anisotropic, related to the affine transformation $A$.

Once the parametric structures of the Gaussian process are defined, the parameters can be estimated via maximum likelihood. 
For the remainder of the article, we adopt a classic regression setting for the trend $\mu(\bs_{\ell};\bbeta) = \bx_{\ell}^{\T} \bbeta$, where $\bx_{\ell}^{\T}$ are rows of the design matrix $\bX$ of dimension $n \times (p+1)$, containing information of a set of $p$ fixed covariates observed at locations $\{\bs_1,\dots,\bs_n\}$. Considering the available information $\bz$, 
estimation of the parameter vector $\bvartheta = (\bbeta^{\T},\bpsi^{\T})^{\T} \in \cR^{p+m+1}$ is given by maximizing the log-likelihood function
\begin{equation}\label{eq:ml}
l(\bvartheta)
= -\frac{n}{2}\log(2\pi) - \frac{1}{2}\log \det\bSigma_{\Ze} - \frac{1}{2}(\bz-\bX \bbeta)^{\T}{\bSigma}_{\Ze}^{-1}(\bz-\bX \bbeta),
\end{equation}
where a vector $\hat{\bvartheta}_{\text{ML}}$ maximizing $l(\cdot)$ is called a Maximum Likelihood estimate (MLe) and is found via numerical optimizers.

Prediction of the process $Z(\cdot)$ at new locations $\{\bs^p_1,\dots,\bs^p_k\}$ are done through the conditional distribution of $Z(\cdot)$ at $\{\bs^p_1,\dots,\bs^p_k\}$ given $\bz$, which follows a multivariate Gaussian distribution defined as
\begin{equation}\label{eq:cond}
    \mathbf{Z}^p | \mathbf{Z} = \bz \sim \mathcal{N}_k(\bX^p\bbeta + \bSigma_{\Pe\Ze}\Sigma_{\Ze}^{-1}(\bz - \bX\bbeta),
    \bSigma_{\Pe} - \bSigma_{\Pe\Ze} \bSigma_{\Ze}^{-1}\bSigma_{\Pe\Ze}^{\T}),
\end{equation}
where $\bSigma_{\Pe\Ze}$ is a matrix of dimension $k \times n$ with elements $[\bSigma_{\Pe\Ze}]_{i,j} = \Cov(\bs^p_i,\bs_j;\bpsi)$, the covariance between the process at unseen locations $\{\bs^p_1,\dots,\bs^p_k\}$, and the process at the observed locations, and $\bSigma_{\Pe}$ is the matrix $k \times k$ with elements $\Cov(\bs_{0_i},\bs_{0_j};\bpsi)$.
Finally, we replace the covariance matrices with the maximum likelihood plug-in estimates in Equation~\eqref{eq:cond}, yielding $\hat{\bSigma}_{\Ze} = \bSigma_{\Ze}(\hat{\bvartheta}_{\text{ML}})$, $\hat{\bSigma}_{\Pe\Ze} = \bSigma_{\Pe\Ze}(\hat{\bvartheta}_{\text{ML}})$, and $\hat{\bSigma}_{\Pe} = \bSigma_{\Pe}(\hat{\bvartheta}_{\text{ML}})$.

In real-world applications, the assumptions of stationarity or isotropy are often times too restrictive, and more flexible covariance functions are needed to ensure the validity of \eqref{eq:ml} and \eqref{eq:cond}. One popular approach to counter the lack of flexibility is the convolution approach introduced by \cite{higdon1999non}, where $Y(\cdot)$ is represented as the convolution of a white noise process $\varphi(\cdot)$, and a spatially-varying smoothing kernel $K(\cdot;\bpsi_{\bs})$ parameterized by a vector $\bpsi_{\bs}$ linked to the spatial location $\bs$, by
\begin{equation*}
    Y(\bs)= \int_{\cD} K\bigl(\bu;\bpsi_{\bs}\bigr)\varphi(\bu)d\bu ,
\end{equation*} \label{eq:2}
leading to the covariance kernel
\begin{equation} \label{eq:c_conv}
\Cov\bigl(\bs_i,\bs_j;\bpsi_{\bs_i},\bpsi_{\bs_j}\bigr) = \int_{\cD}K\bigl(\bu;\bpsi_{\bs_i}\bigr)K\bigl(\bu;\bpsi_{\bs_j}\bigr)d\bu .
\end{equation}
The requirement on the kernel function is simply that $\int_{\cR^d}K^d(\bu;\bpsi_{\bs})d\bu < \infty$, leading to positive definite covariance functions. As opposed to defining nonstationary covariance functions directly, we can obtain valid nonstationary covariance functions by simply defining valid kernels, making the convolution approach more appealing when modeling nonstationary processes in the covariance.
\cite{paciorek2006spatial} introduced a class of models based on \eqref{eq:c_conv} for which the integrations can be carried out analytically. 
They define the smoothing kernels $K_{\ell}(\cdot) = K(\cdot;\bpsi_{\bs_{\ell}})$ as multivariate Gaussian kernels centered at location $\bs_{\ell}$, leading to a covariance function with an integration-free form. The resulting covariance function can also be extended considering a spatially-varying smoothness \citep{stein2005nonstationary} as well as nonstationarity in the variance, leading to the nonstationary covariance function
\begin{equation} \label{eq:c_ns}
    \Cov_{NS}(\bs_i,\bs_j; \bpsi_i,\bpsi_j)= \sigma_i \sigma_j |\bSigma_i|^{1/4}|\bSigma_j|^{1/4} \Biggl|\frac{\bSigma_i + \bSigma_j}{2}\Biggr|^{-\frac{1}{2}} \cM_{(\nu_i + \nu_j)/2}\Bigl(\sqrt{Q_{ij}}\Bigr) ,
\end{equation}
where $\sigma_{\ell} = \sigma(\bs_{\ell})$ is a standard deviation process, $\nu_{\ell} = \nu(\bs_{\ell})$ is a smoothness process, $\cM_{\nu}(\cdot)$ is the Mat\'ern correlation function with smoothness $\nu$ and a deliberate valid scale parameter, $\bSigma_{\ell} = \bSigma(\bpsi_{\ell})$ is a $2\times2$ positive-definite covariance matrix process of the Gaussian kernel (i.e., the covariance kernel), and where $Q_{ij} = h_{(\bSigma_{i} + \bSigma{j})/2}$ is a semi-metric distance function \citep{schoenberg1938metric} defined as
\begin{equation}\label{eq:qij}
    Q_{ij}=(\bs_i - \bs_j)^{\T}\Bigl(\frac{\bSigma_i + \bSigma_j}{2}\Bigr)^{-1}(\bs_i - \bs_j) , \ \ \bs_i , \bs_j \in \cD ,
\end{equation}
mimicking a geometrical anisotropic distance with an affine matrix defined by the average of the two covariance kernels at locations $\bs_i$ and $\bs_j$.
Based on $\Cov_{NS}(\cdot,\cdot)$, \citealt{risser2015regression} adopts parametric functions for the spatial standard deviation $\sigma(\cdot)$ and the local anisotropic structure $\bSigma(\cdot)$, seeking a low-dimensional parametric space, stating a compromise between the flexibility of $\Cov_{NS}(\cdot,\cdot)$ and computational requirements. 
They assume a linear model for the logarithm of the standard deviation, while for the covariance kernel, they follow the parametric model in \citealt{hoff2012covariance} defined as
\begin{equation*}
    \bSigma(\bs_{\ell}) = \mathbf{A} + \mathbf{B} \bx_{\ell}\bx_{\ell}^{\T}\mathbf{B}^{\T},
\end{equation*}
where $\mathbf{A}$ is a $d\times d$ symmetric positive definite matrix representing an error covariance and where $\mathbf{B}$ is a $d \times p$ matrix of rank $1$ with coefficients describing how additional variability is distributed across the $d$ dimensions. They comment that $\mathbf{A}$ is identifiable and $\mathbf{B}$ is identifiable up to a sign, given an appropriate range of covariate values \citep{hoff2012covariance}.

Although these covariance functions are able to represent nonstationary processes, they come with certain limitations.
The nonparametric nature of $\Cov_{NS}(\cdot;\cdot)$ is prone to numerical and computational difficulties, as well as at risk of overfitting.
On the other hand, the rank one anisotropic matrix model used by Risser and Calder imposes extra assumptions over the covariance kernel, leading to the off-diagonal elements of $\bSigma(\cdot)$ being modeled with the same parameters as the diagonal elements. A full-rank model can overcome this limitation but at a costly increase in the total number of parameters.
Furthermore, although a spatially-varying parametric function for the smoothness process is introduced theoretically, the implementation later reverts to a global smoothness parameter.

\section{Modular nonstationary covariance functions}\label{ch:consider_covs}
This section introduces a class of covariance functions based on the covariate regression framework, designed to offer a general-purpose covariate function capable of providing a convenient tradeoff between flexibility and computational efficiency. We achieve this by defining a set of parametric spatially-varying functions for various sources of nonstationarity represented by \eqref{eq:c_ns}, employing a frequentist approach to benefit from scalable and streamlined frameworks. We begin by presenting the model in Section~\ref{sec:dense_model}, then we explore the interpretability of the model in Section~\ref{sec:interpretability}, and discuss how this covariance function can be adapted to handle very large datasets in Section~\ref{sec:sparse_model}. We conclude by introducing some challenges and strategies for regularization and model selection in Section~\ref{sec:challenges} related to covariate-based covariance functions.

\subsection{A class of dense nonstationary covariate-based covariance functions}\label{sec:dense_model}
To create an economical yet flexible class of nonstationary covariance functions, we assume first that the underlying smooth spatial structure $Y(\cdot)$ follows the class of covariance functions given in \eqref{eq:c_ns}, but by considering parametric spatially-varying functions for $\sigma(\cdot), \nu(\cdot)$, and $\bSigma(\cdot)$ instead as stochastic processes.
These structures define the multivariate function $\bPsi(\cdot)$ (likely nonparametric in nature), which is based on a set of fixed and observable covariates $\bx_{\ell}^{*}$ at a given location $\bs_{\ell}$, yielding $\bpsi_{\ell} = (\sigma_{\ell},\text{vech}{(\bSigma_{\ell}),\nu_{\ell}})^{\T}$, where $\text{vech}(\cdot)$ vectorizes the upper half of a $d\times d$ symmetric matrix into vector of length $d(d+1)/2$. 
In practice, we approximate $\bPsi(\cdot)$ with a parametric surrogate $\tilde{\bPsi}(\cdot;\bphi)$, where each component is driven by a small set of covariates $\bx_{\ell} \in \mathcal{R}^{p+1}$ (with $p \ll w$), and a low-dimensional parameter vector $\bphi$.
We restrict each component of $\tilde{\bPsi}(\cdot;\bphi)$ to smooth functions, as well as employ functions that are computationally efficient to evaluate. By requiring the components of $\tilde{\bPsi}(\cdot;\bphi)$ to evolve smoothly, we ensure that nearby locations exhibit similar covariance structure, thereby coherently linking local process properties as in the nonstationary Mat\'ern construction of \citep{paciorek2006spatial}.

Among the considered sources of nonstationarity, the anisotropic structure $\bSigma$ is one of the most challenging to model since in the spatial domain, $\bSigma(\cdot)$ yields positive definite, symmetric $2\times2$ matrices, with three unique elements, $\bSigmap{1}{1},\bSigmap{2}{2}$, and $\bSigmap{1}{2}$.
The function we propose for $\bSigma(\cdot;\btheta)$ shapes the size of the kernel in each axis, with a third component redistributing the trace of the kernel matrix, contributing to the tilt of the covariance kernel.
We propose the following models for each of the elements of $\bSigma$
\begin{equation}
    \bSigma(\cdot;\btheta) = \rho(\cdot;\btheta_{ms})^2 \begin{pmatrix}
        1 & r(\cdot;\btheta_{ga}) \cos(\omega(\cdot;\btheta_{tt})) \\
          & r(\cdot;\btheta_{ga})^2
    \end{pmatrix},
\end{equation}
where $\rho(\bx_{\ell};\btheta_{ms}) = \exp(\bx_{\ell}^{\T} \btheta_{ms})$ governs the size of the kernel, controlled by a parameter vector $\btheta_{ms}$, $r(\bx_{\ell};\btheta_{ga}) = \exp(\bx_{\ell}^{\T} \btheta_{ga})$ controls the shrink or expansion of the kernel in the secondary axis, controlled by a parameter vector $\btheta_{ga}$, and where $\cos(\omega(\bx_{\ell};\btheta_{tt})) = \cos(\text{logit}^{-1}(\bx_{\ell}^{\T} \btheta_{tt}) \pi)$ indirectly controlling the rotation of the anisotropic matrix, controlled by a parameter vector $\btheta_{tt}$. 
We consider all the parameters handling the anisotropic structure in a vector $\btheta = (\btheta_{ms}^{\T},\btheta_{ga}^{\T},\btheta_{tt}^{\T})^{\T}$. The subscripts $ms$, $ga$, and $tt$ relate to main scale, geometric anisotropy, and tilt, i.e., sources of nonstationarity of the local geometric anisotropy $\bSigma(\cdot;\btheta)$.
This specific parameterization is similar to that derived from modeling the spherical coordinates of the Cholesky factor of $\bSigma$, rather than its raw elements \citep{pinheiro1996unconstrained}. 

Continuing with the functions of $\tilde{\bPsi}(\cdot;\bphi)$, we have that $\text{Var}(Y(\bs_{\ell}))=\sigma^2(\bs_{\ell})$. 
We adopt the following model for the marginal standard deviation $\sigma(\cdot)$
\begin{equation*}
    \sigma(\bx_{\ell};\balpha) = \exp\left(0.5 \bx_{\ell}^{\T} \balpha\right),
\end{equation*}
where $\balpha$ is the associated parameter vector. 
However, the joint estimation of $\balpha$ and $\btheta_{ms}$ will often lead to a set of highly correlated pairs of parameters associated to the same covariate (including the intercept), leading to an almost perfect correlation for very large sample sizes, as also mentioned in \citep{paciorek2003nonstationary}. 
To ease these correlations while being able to retain direct interpretability, we consider instead the parameters $\balpha^{(d)}$ and $\btheta^{(d)}_{ms}$ as $\balpha^{(d)} = \balpha + \btheta_{ms}$, and $\btheta^{(d)}_{ms} = \balpha - \btheta_{ms}$, when a specific covariate $X_{\ell}$ is considered in both the standard deviation and the spatial scale functions.


Unlike the variance of the process, the smoothness parameter $\nu$ is frequently fixed at half-integer values based on expert judgment, a practice part of the geostatistical folklore \citep{de2022information}. This practice arises from theoretical and numerical challenges inherent to the Mat\'ern covariance family, where fixing $\nu$ to half-integers simplifies computation and reduces numerical instabilities. However, from a theoretical standpoint, $\nu$ is microergodic under infill asymptotics (\citealt{karvonen2023asymptotic},  \citealt{stein1999interpolation}, Section 6.2), meaning it can be consistently estimated despite the non-identifiability of non-microergodic parameters (e.g., variance and scale) in bounded domains. 
Numerically, estimating $\nu$ beyond half-integers remains challenging due to costly Bessel function evaluations $\mathcal{K}_\nu(\cdot)$ \citep{chen2024identifiability}. While extending $\nu$ to vary spatially offers significant advantages, particularly in capturing localized roughness variations (e.g., abrupt transitions between smooth and rough regions), it exacerbates computational instability, making unconstrained models like \eqref{eq:c_ns} notoriously difficult to fit \citep{stein2005nonstationary}.

To reconcile these competing demands, we propose a parametric model for the spatially-varying smoothness that restricts the range of variability and consider a numerically more stable approach when combining smoothness between locations. We implement the latter by, as opposed to modeling the spatially-varying smoothness as $(\nu_i +\nu_j)/2$, representing it as $\sqrt{\nu_i\nu_j}$, defining fundamentally different interweaving behaviors. 
While both representations behave similarly when $\nu_i \approx \nu_j$, $\sqrt{\nu_i\nu_j}$ yields a more conservative interweaving under highly discrepant smoothness. This leads to numerically more stable covariance matrices when compared to those based on $(\nu_i +\nu_j)/2$, allowing the representation of processes with highly heterogeneous smoothness at a local level, which are suitable, for example, for modeling strong discrepancies in the data, such as sharp jumps in the process.
Moreover, to reduce the numerical challenges associated with the estimation of extreme values of smoothness, which can be of little relevance in most applications, we model $\nu_i$ such that it restricts the range of variation, adding another layer of numerical stability. We constrain its variability within specified lower and upper bounds to better capture the inherent smoothness of the spatial process, leading to
\begin{equation} \label{eq:spatially_var_smtns}
    \nu(\bx_{\ell};\bzeta) = \frac{\numax - \numin}{1+\exp(-  \bx_{\ell}^{\T} \bzeta)} + \numin,
\end{equation}
where $\bxi$ is the associated parametric vector, and where $0 < \numin \leq \numax < \infty$ refers to the lower and upper bounds of the smoothness.
The proposed model \eqref{eq:spatially_var_smtns} relates to a shifted logistic cumulative density function with a fixed scale parameter, where we focus on shaping its location parameter.

This leads to the following general modular regression-based covariance function
\begin{equation} \label{eq:gen_rns_ours}
    \Cov_{\text{GR}}(\bs_i,\bs_j;\bx_i,\bx_j,\bphi)= \sigma(\bx_i;\balpha) \sigma(\bx_j;\balpha) \frac{|\bSigma(\bx_i;\btheta)|^{1/4}|\bSigma(\bx_j;\btheta)|^{1/4}} {
        \left|\frac{\bSigma(\bx_i;\btheta) + \bSigma(\bx_j;\btheta)}{2}\right|^{{1}/{2}}
     } \mathcal{M}_{\sqrt{\nu(\bx_i;\bxi)\nu(\bx_j;\bxi)}}\Bigl(\sqrt{Q_{ij}}\Bigr),
\end{equation}
where $\bphi = (\balpha^{\T},\btheta_{ms}^{\T},\btheta_{ga}^{\T},\btheta_{tt}^{\T},\bxi^{\T})^{\T}$. In Appendix~\ref{ap:proof} we show that the resulting covariance function, particularly with the proposed parametric spatially-varying function for the smoothness, is positive definite.

\subsection{Interpretability} \label{sec:interpretability}

The presented covariance model \eqref{eq:gen_rns_ours} provides a closed-form expression of the second-order structure of $Y(\cdot)$, by interweaving locally stationary geometrically anisotropic structures. In a small neighborhood around $\bs_{\ell}$, $ \bs_i \approx \bs_{\ell}$, Equation~\eqref{eq:gen_rns_ours} simplifies to
\begin{equation} \label{eq:local_matern}
    \Cov_{GR}(\bs_i,\bs_{\ell};\bx_{\ell},\bphi) \approx \sigma(\bx_{\ell};\balpha)^2  \cM_{\nu(\bx_{\ell};\bxi)}\Bigl(\sqrt{(\bs_{\ell} - \bs_i)^{\T} \bSigma(\bx_{\ell};\btheta)^{-1} (\bs_{\ell} - \bs_i)}\Bigr),
\end{equation}
defining stationary Mat\'ern covariance function with geometrically anisotropic matrix $\bSigma(\bx_{\ell};\btheta)$, smoothness $\nu_{\ell}$ and variance $\sigma(\bx_{\ell};\balpha)^2$. While $\bphi$ unveils how different covariates influence a specific source of nonstationarity, inspecting the parametric structure at $\bx_{\ell}$ unveils its local dependency structure.
Parameters linked with covariates are centered, implying no effect due to the covariate $\bx_{\ell}$ is present over the specific spatially-varying function when $\btheta_{\ell} = 0$.
Straightforward derivation of these relationships is that unit increases in $\bx_{\ell}$, holding other variables constant, results in the overall variance multiplying by $\exp(\balpha_{,i})$, while for the smoothness parameters $\bxi$, the inverse of the logit transformation of a parameter is associated with the local smoothness of the process.

Considering now the local anisotropy structure, the function $\bSigma$ defines a $2\times2$ symmetric positive definite matrix, which relates to ellipsoids of specific shape. 
Since we provide a model directly for the elements of $\bSigma$, it is not clear how the parameters of $\btheta$ shape the associated kernel, for example, compared to a more classical representation of the kernel in spatial statistics as the spectral representation of $\bSigma$. To reveal this, we retrieve the associated eigenvalues and eigenvectors. 
The eigenvalues of the associated $2\times2$ symmetric positive definite matrix are defined by
\begin{align*}
    e_{\ell,i} = \frac{\rho^2_{\ell}}{2} \Bigg[(r^2_{\ell} + 1) - (-1)^{i} \sqrt{(r^2_{\ell}+1)^2 - 4r^2_{\ell}\sin(\omega_{\ell})^2} \Bigg], i = 1,2 .
\end{align*}
In general, the eigenvalues of $\bSigma$ are well defined, except for the limiting cases when $\omega = 0$ or $\omega= \pi$, where one of the eigenvalues collapses, though this is not a practical concern.
As special cases, the eigenvalues simplify to $(\rho^2_{\ell},\rho^2_{\ell} r_{\ell})$ when $\omega_{\ell}=\pi/2$, and to $\bigl(\rho^2_{\ell}\bigl(1 + \cos(\omega_{\ell})\bigr), \rho^2_{\ell}\bigl(1 - \cos(\omega_{\ell})\bigr)\bigr)$ when $r_{\ell}=1$. 
Under these scenarios, both $r$ and $\omega$ redistribute the trace of $\bSigma$ along each eigenvalue, with the particularity that $r_{\ell}$ also influences the trace of $\bSigma$, while $\omega_{\ell}$ does not. 

\begin{figure}[!h]
    \centering
    \begin{subfigure}[c]{0.33\textwidth}
    \centering
    \includegraphics[keepaspectratio,height=3cm]{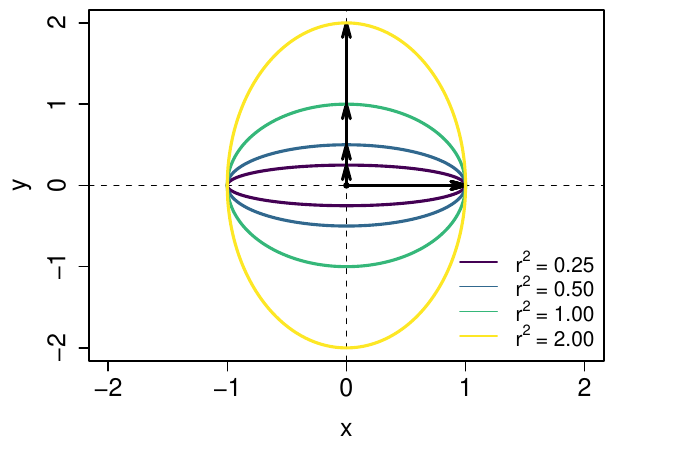}
    \caption{}
    \end{subfigure}
    \begin{subfigure}[c]{0.33\textwidth}
    \centering
    \includegraphics[keepaspectratio,height=3cm]{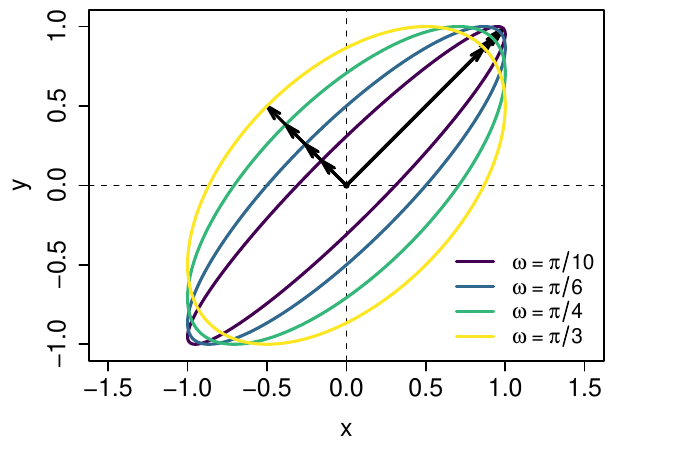}
    \caption{}
    \end{subfigure}
    \begin{subfigure}[c]{0.33\textwidth}
    \centering
    \includegraphics[keepaspectratio,height=3cm]{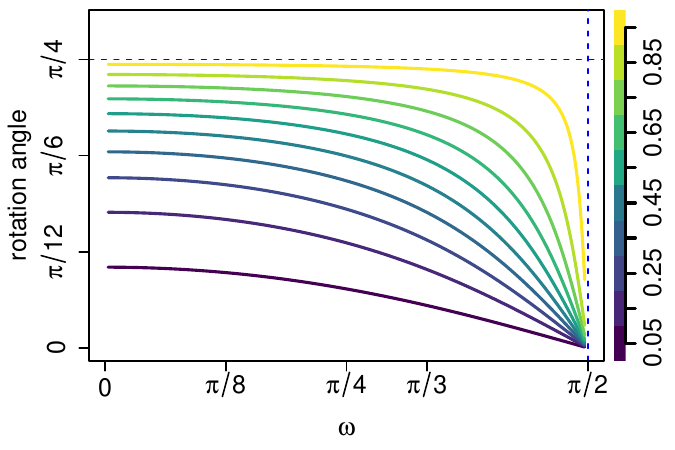}
    \caption{}
    \end{subfigure}
    \caption{Kernel ellipses defined by the eigenvalues with univariate changes of $r^2$ (a) and $\omega$ (b), when  $\omega = \pi/2$ and $r=1$, respectively. In Panel~(c) rotation angle (in radians) of the associate ellipsoid for different values of $r<1$ (colored lines) and $\omega \leq \pi/2$.}
    \label{fig:univariate_variations}
\end{figure}

\begin{figure}[!h]
    \centering
    \begin{subfigure}[c]{0.33\textwidth}
    \centering
    \includegraphics[keepaspectratio,height=3cm]{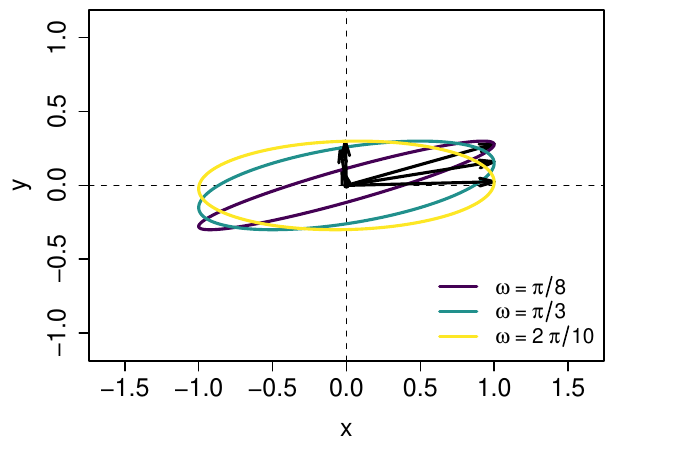}
    \caption{}
    \label{fig:cat_a}
    \end{subfigure}
    \begin{subfigure}[c]{0.33\textwidth}
    \centering
    \includegraphics[keepaspectratio,height=3cm]{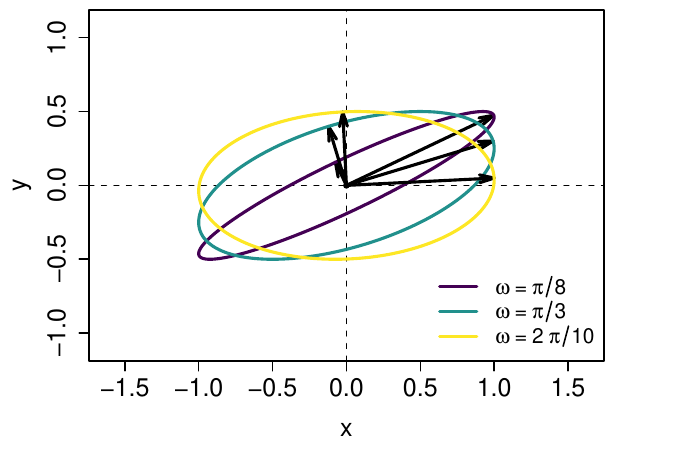}
    \caption{}
    \end{subfigure}
    \begin{subfigure}[c]{0.33\textwidth}
    \centering
    \includegraphics[keepaspectratio,height=3cm]{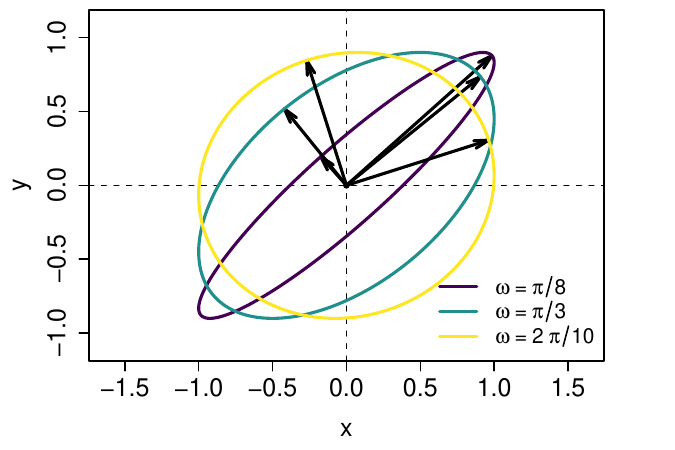}
    \caption{}
    \end{subfigure}
    \caption{Ellipses defined by the eigvenvalues for different values of $\omega$ and $r$. In Panel~(a) $r^2 = 0.3$, Panel~(b) $r^2 = 0.5$, and $r^2=0.9$ in Panel~(c).}
    \label{fig:bivariate_alpha_r}
\end{figure}

To further inspect the representation of the ellipsoid based on $\bSigma$, we retrieve the associated eigenvectors. Each eigenvalue is associated with a specific eigenvector of the form
\begin{align*}
    \mathbf{e}_{\ell,i} = \frac{s}{\sqrt{\Bigl(2r_{\ell} \cos(\omega_{\ell})\Bigr)^2 + \Bigl(r^2_{\ell} - 1 - (-1)^i A_{\ell}\Bigr)^2}} \begin{pmatrix}
        2r_{\ell} \cos(\omega_{\ell}) \\
        r^2_{\ell} - 1 - (-1)^i A_{\ell}
    \end{pmatrix}, i = 1,2,
\end{align*}
where $A_{\ell} = \sqrt{(r^2_{\ell}+1)^2 - 4r^2_{\ell}\sin{(\omega_{\ell})}^2}$. The rotation angle of the anisotropic matrix is given by $\arctan{\Big(\frac{r^2_{\ell} - 1 + A_{\ell}}{2r \cos(\omega_{\ell})}\Big)}$.
In Figure~\ref{fig:univariate_variations} we present kernel ellipses defined by the eigenvalues with univariate changes with respect to $r^2$ in Panel~(a), and to $\omega$ in Panel~(b). In Panel~(c), we show the rotation angle (in radians) of the associate ellipsoid for different values of $r^2$ and $\omega$, for values of $r^2 \leq 1$, and $\omega \leq \pi/2$. While the rotation angle is mainly driven for values of $\omega$ near $\pi/2$, the maximum rotation angle is restricted by $r^2$,  which mainly controls the rotation angle for low values of $\omega$. 
Then, we can relate $\omega$ with the limited rotation of the ellipsoid. This behavior can be seen in more detail in Figure~\ref{fig:bivariate_alpha_r}, where bivariate changes of $\omega$ and $r^2$ are presented. Under bivariate changes, expansion of the kernel in $y$-axis are given by $\sqrt{\rho r} \sin{(\omega)}$, and $\sqrt{\rho} \sin{(\omega)}$ in the $x$ axis.

How the kernel $\bSigma$ behaves is directly related to the representation of $\sqrt{Q}$ locally.
The proposed function for $\bSigma(\cdot;\btheta)$ generalizes the common global types of anisotropic structures. In its simplest form, $\btheta_{ga} = \btheta_{tt} = 0$, and $\btheta_{ms} = \theta_{ms}$, simplifies to a global isotropic scenario with a scale parameter equal to $\exp(\theta_{ms})$.
Global anisotropic structures can be retrieve when $\btheta_{ga} = \theta_{ga} \neq 0$ or when $\btheta_{tt} = \theta_{tt} \neq 0$.
Considering $\btheta_{tt} = 0$, and for locations $\bs_i$ and $\bs_j$ in a small neighborhood around a location $\bs_{\ell}$ the scale for a given $\bs_i - \bs_j = (\Delta_x, \Delta_y)^\T$ is then given by
\begin{align*}
   \sqrt{Q_{ij}} = \rho_{\ell}^{-1}\sqrt{\frac{r_{\ell}^2 \Delta_x^2 + \Delta_y^2}{r^2_{\ell}}},
\end{align*}
where pure differences in the $x$-axis are related to a scale of $\rho_{\ell}$, while pure differences in the $y$-axis have associated a scale that shrinks or expands by the parameter $r_{\ell}$. $r_{\ell}$ takes the role of that associated with geometric anisotropy from classical geostatistics.
Once we also consider $\btheta_{tt} \neq \mathbf{0}$, the semi-distance metric takes the shape of
\begin{align*}
   \sqrt{Q_{ij}} = \rho_{\ell}^{-1} \sqrt{\frac{r_{\ell}^2 \Delta_x^2 + \Delta_y^2 + 2\Delta_x\Delta_y r_{\ell} \cos{(\alpha_{\ell})}}{r^2_{\ell} \sin^2{(\alpha_{\ell})}}},
\end{align*}
where pure differences in the $x$-axis will lead to a scale given by $\rho_{\ell} \sin{(\alpha_{\ell})}$. On the other hand, pure differences in the $y$-axis will lead to a scale given by $\rho_{\ell} r_{\ell} \sin{(\alpha_{\ell})}$.

Based on the eigen-decomposition of the local anisotropy structure, we have seen that the parameters $\rho$, $r$, and the restricted tilt collectively determine $\bSigma$, so their effects are inherently entangled in shaping the kernel. However, despite this geometric interdependence, the modular construction assigns separate parametric functions to each component of anisotropy where each aspect ($\rho$, $r$, and restricted tilt) is governed by its own set of covariate coefficients ($\btheta_{ms}, \btheta_{ga}, \btheta_{tt}$, respectively). 
This means that the associated parameter vectors remain structurally distinct and interpretable. 
Based on this modular design, one can pursue different strategies for regularization (e.g. to protect against overfitting or improve numerical stability). One can impose a single, global penalty or prior on a functional of the full anisotropy matrix (for example its determinant or condition number), or apply separate penalties or priors directly to each parameter block (benefiting from their unconstrained, centered formulation), thereby controlling scale, ratio, and restricted tilt independently for more granular control and streamlined implementation. In Section \ref{sec:challenges} we follow the second approach by introducing a penalty on the global scale parameter to stabilize the covariance function, followed by penalties over the covariate-driven parameters in each $\btheta$ for individualized regularization.


The presented parameterization for the anisotropic matrix differs from the standard spectral decomposition $\bSigma(\btheta) = \mathbf{R}(\btheta) \text{diag}(e_1, e_2) \mathbf{R}^\top(\btheta)$. When $e_1 \approx e_2$, rotational unidentifiability emerges due to a flattened likelihood surface, as small perturbations in $\btheta$ yield negligible changes in $\bSigma(\cdot)$. This near-isotropic scenario also induces a high correlation between eigenvalues, complicating inference. Moreover, under anisotropy ($e_1 \neq e_2$), the spectral representation exhibits sensitivity: small angular adjustments disproportionately alter the matrix structure, creating ridges or plateaus that can affect numerical optimization.
To cope with these limitations, our model encodes orientation implicitly as a by-product of redistributing the trace of the anisotropic matrix through a dominant isotropic component and a secondary anisotropic term. This avoids explicit angle specification, reducing sensitivity to directional fluctuations. Moreover, by defining anisotropy relative to the dominant scale, we mitigate parameter correlations and ensure full rank. Unlike \citet{risser2015regression}, our modular structure enables targeted control of scale, geometric anisotropy, and orientation without imposing specific assumptions over the anisotropic matrix.

\subsection{Inducing sparsness over the covariate-based covariance function} \label{sec:sparse_model}
Employing dense covariance matrices such as \eqref{eq:gen_rns_ours} can be challenging for large sample sizes, mainly due to the evaluation of $\det(\bSigma_{\Ze})$ and solving linear systems involving $\bSigma_{\Ze}$, requiring $O(n^3)$ floating point operations and $O(n^2)$ memory.
A well-known workaround for computational challenges in spatial prediction is the Tapering Approach (TA) \citep{furrer2006covariance}, which aims to induce sparsity into the covariance matrix to benefit from fast and reliable algorithms.
The sparsity is achieved by multiplying element-wise the covariance matrix $\bSigma_{\Ze}$ with a valid compact-supported correlation matrix, known as the taper matrix.

Considering $\mathbf{T}_{\delta}$ as the $n \times n$ positive-definite taper matrix based on a compact-supported correlation function with scale parameter $\delta$, the tapered matrix is then defined as 
\begin{equation}\label{eq:tapering}
 \bSigma_{\Te} = \bSigma_{\Ze} \odot \mathbf{T}_{\delta},
\end{equation}
where $\odot$ denotes the Schur or element-wise product, and $\delta$ controls the number of induced zeroes in the tapered matrix. 
When $\delta \to 0$, $\bSigma_{\Te}$ simplifies to a diagonal matrix, and when $\delta \to \infty$, recovers $\bSigma_{\Ze}$. 
In practice, strong sparsity can be induced, often leading to a covariance matrix with only $1\%$ of non-zero elements. As a reference, for stationary models such as the Mat\'ern, the taper scale parameter can be chosen such that we include between $50$ to $100$ neighbors should be sufficient for reliable predictions \citep{blasi2022selective}. Another strategy is to consider selection via cross-validation or by matching the process's effective scale \citep{furrer2006covariance}. 

The presented covariance model \eqref{eq:c_ns} can easily be adapted for TA.
We propose a simplification of the nonstationary model \eqref{eq:gen_rns_ours} by considering only spatially-varying local structures concerning scale, smoothness, and variance, leading to
\begin{equation}\label{eq:taper_function}
 \Cov_{\text{s}}(\bs_i,\bs_j;\bx_i,\bx_j,\bphi)= \sigma_i \sigma_j 2 \frac{\rho_{i}^{1/2} \rho_{j}^{1/2}}{\rho_i + \rho_j} \mathcal{M}_{\sqrt{\nu(\bx_i;\bxi)\nu(\bx_j;\bxi)}}\Biggl(\sqrt{\frac{h}{\frac{\rho_i + \rho_j}{2}}}\Biggr),
\end{equation}
where $Q_{ij}$ has simplified to $h/\frac{\rho_i + \rho_j}{2}$, which entails a local stationary structure at location $\bs_{\ell}$ with a scale parameter $\rho_{\ell}$. Given that this is a special case to the family of covariance functions presented in \ref{ch:consider_covs}, it is still positive definite.
The considered sources of nonstationarity yield a set of local stationary isotropic spatial structures that differ in smoothness, variance, and scale.
The proposed model for the spatially-varying smoothness presents two advantages under this framework. 
Firstly, under very large sample sizes, it is convenient to control the extent of variability of the spatially-varying smoothness, which can cause numerical instabilities. 
Secondly, under TA, it is common practice to select a common taper function that is at least as smooth as the dense covariance function, which can be achieved by the $\numax$ hyperparameter.
In Appendix~\ref{sec:A.taper_covs}, we present some of the most popular compact-supported correlation functions, such as the Spherical and Wendland covariance models \citep{bevilacqua2019estimation}.

Estimation of $\bvartheta$ is done by maximizing
\begin{equation}\label{eq:taper_loglik}
    l_{\Te}(\bvartheta)
= -\frac{n}{2} \log(2\pi) - \frac{1}{2}\log \det(\bSigma_{\Ze} \odot \Te_{\delta}) - \frac{1}{2}(\bz-\bX \bbeta)^{\T}({\bSigma_{\Ze} \odot \Te}_{\delta})^{-1}(\bz-\bX \bbeta).
\end{equation}
In practice, there is a tradeoff between the taper scale $\delta$ and the bias of the estimates, especially if $\delta$ is small relative to the true correlation scale of the process. 

\subsection{Challenges of covariate-based covariance functions} \label{sec:challenges}

The covariate-based covariance model offers a good balance between flexibility and computational efficiency while adding a potential layer of interpretability, depending on the sampling scheme. 
However, certain types of covariates may distort the local properties of the spatial structure, and often times regularization is required, as is common to nonstationary covariance functions.

A key assumption of convolution-based covariate functions is that the kernels evolve smoothly over the study domain $\cD$. This allows to link local properties of the spatial structure of the process with the functional form of the proposed covariance function. The validity of this assumption is sensible to the functional form of $\tilde{\bPsi}(\cdot;\bphi)$ and consequently depends on the nature of the covariates used. Thus, it naturally raises questions about the types of covariates suitable for modeling $\tilde{\bPsi}(\cdot;\bphi)$. For example, ordinal and noisy covariates will not meet this assumption and will impose spurious behaviors. Given that these types of covariates are frequently available when modeling spatial data, we comment on the consequences of employing these types of covariates over the covariance function.

\begin{figure}[!ht]
    \centering
    \begin{subfigure}[c]{0.33\textwidth}
    \centering
    \includegraphics[keepaspectratio, height=3cm]{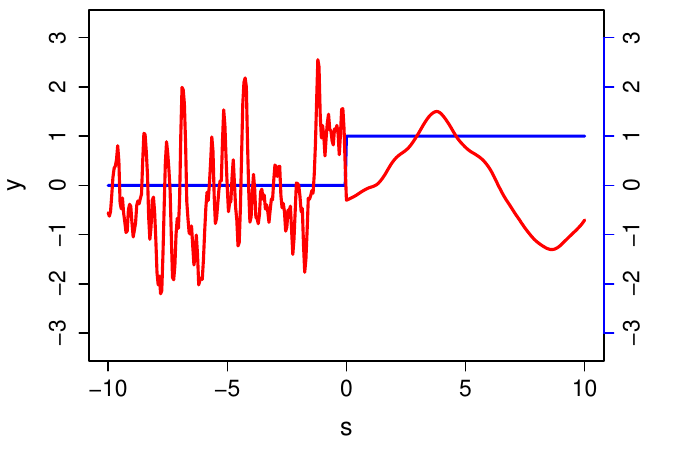}
    \caption{}
    \end{subfigure}
    \begin{subfigure}[c]{0.33\textwidth}
    \centering
    \includegraphics[keepaspectratio,height=3cm]{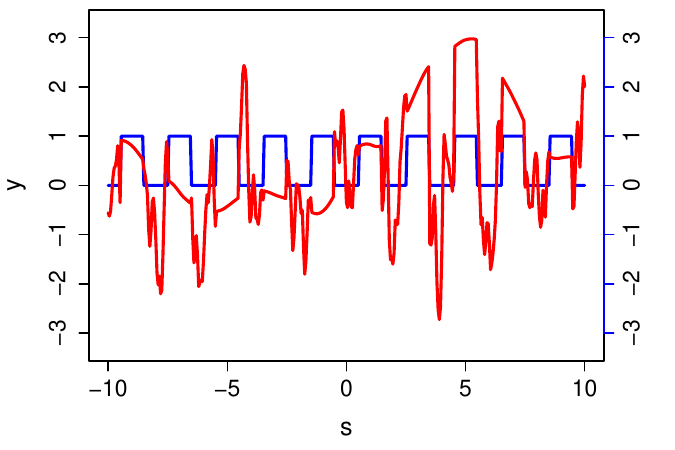}
    \caption{}
    \end{subfigure}
    \begin{subfigure}[c]{0.33\textwidth}
    \centering
    \includegraphics[keepaspectratio,height=3cm]{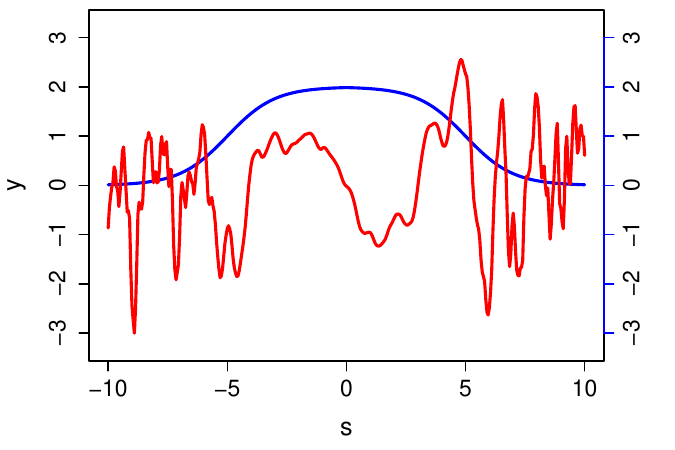}
    \caption{}
    \end{subfigure}
    \begin{subfigure}[c]{0.33\textwidth}
    \centering
    \includegraphics[keepaspectratio,height=3cm]{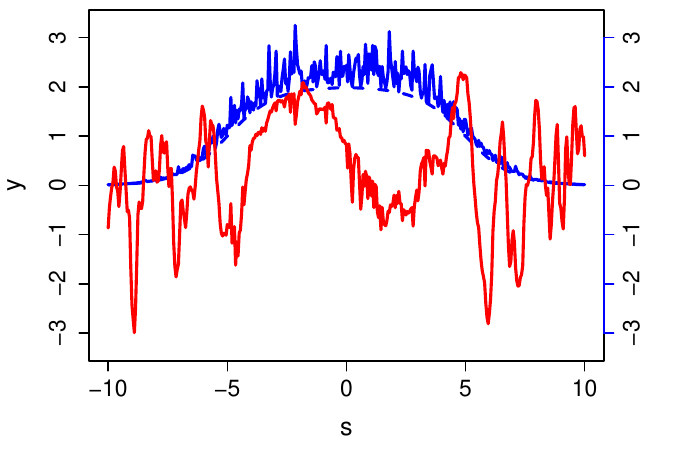}
    \caption{}
    \end{subfigure}
    \caption{In red, realizations of stationary Gaussian processes in $\mathcal{R}^1$. In the first two Panels, the spatially-varying scale with an ordinal variable is modeled (with one jump and several jumps). The last two Panels, with a smooth and noisy covariate, respectively. In blue is the covariate information. All process realizations were simulated with the same seed.}
    \label{fig:covariates_problem}
\end{figure}

When considering ordinal variables, while the local stationarity assumption holds within regions sharing the same level, covariance between levels can exhibit a reduced correlation regardless of the spatial scale. 
In Panel~(a) and Panel~(b) of Figure~\ref{fig:covariates_problem} we present an example of a process realization in $\cR^1$. 
In $\mathcal{R}^1$, the anisotropic structure simplifies to $\bSigma(\cdot) = \rho(\cdot)$, a varying scale function in $\mathcal{R}^1$.
Although the local stationarity assumption holds for values of the ordinal variable of the same level, there is a break between levels. The reduction of correlation between levels is caused by what is known as the prefactor, i.e., the ratio between the product of determinants in the dividend, and the determinant in the divisor, as shown in \eqref{eq:c_ns}. This is due to the fact that when one anisotropic matrix has a much larger determinant than the other, the denominator grows faster than the numerator, which causes the overall prefactor, and hence the correlation, to be smaller than it would be if the two determinants were equal. This means that two locations both characterized by weak determinants may exhibit a stronger mutual correlation than a pair in which one location has a weak determinant and the other a strong one.
Moreover, this spurious effect is present regardless of how close and strong the spatial scale is at both levels. 
For example, assuming local scales of $1$ and $10$ at each level would lead to a prefactor of approximately $0.6$, meaning that irrespective of how close the locations between levels are, the maximum achievable correlation between regions will be of $0.6$. 

The same spurious behavior is also present when employing noisy covariates, specifically, when noisy covariates are considered in the covariance kernel $\bSigma(\cdot;\btheta)$ and the smoothness $\nu(\cdot,\cdot;\bxi)$. The former aligns when employing categorical variables, inducing an overall reduction of the correlation due to the prefactor, while employing noisy covariates in the smoothness can trigger numerical challenges due to the discrepancy of local behaviors, impacting the near diagonal elements of the covariance matrix. 
In Figure~\ref{fig:covariates_problem} we exemplify these behaviors in $\mathcal{R}^1$. We see that although in Panel~(d), the associate scale should increase given that the noisy version is at least the value of the smooth covariate (implying a stronger correlated process), the noise of the covariate leads to a lower than expected correlated process.

The last challenge we address in this section concerns numerical stability and model selection. For Matern-like covariance matrices under single realization on a fixed domain, parameters such as the scale and variance cannot be separately and consistently estimated, only the combined microergodic parameter $\sigma^2 \rho ^{2\nu}$ is identifiable \citep{zhang2004inconsistent}. In practice, this manifests as a likelihood ridge along which different combinations of $\sigma^2$ and $\rho$ yield almost identical fits, with a greater risk of resulting in ill-conditioned optimization. Because our model builds from the Mat\'ern framework, strategies to cope with these types of drawbacks are also required. To reduce this, we introduce a penalty over the baseline smoothness-scale product. Specifically, we penalize the product $\sqrt{\nu_0} \rho_0$, where $\nu_0$ and $\rho_0$ denote the smoothness and scale parameters when all $\bx_\ell = 0$, for $\ell > 1$. This yields the penalized likelihood function:
\begin{equation}\label{eq:pen}
l_{pen}(\bvartheta) = l(\bvartheta) + n \lambda_r \sqrt{\nu_0} \rho_0,
\end{equation}
where $\lambda_r \geq 0$ is the regularization parameter and $\hat{\bvartheta}_{pen}$ is the maximizer of $l_{pen}(\cdot)$. By penalizing $\sqrt{\nu_0} \rho_0$, we discourage highly smoothed, long-tailed covariance functions, which are particularly challenging from a numerical perspective. When $\lambda_r = 0$, the method reduces to standard maximum likelihood. As $\lambda_r \to \infty$, the penalty enforces $\rho_0 \to 0$ and $\nu_0 \to \nu_{\min}$, simplifying the covariance to its least smooth, shortest-scale form.
Rather than adding a nugget effect, which creates a discontinuity at the origin, our method maintains exact interpolation at all observed sites, which is a desirable property in deterministic application such as emulation or simulation-based models \citep{peng2014choice}. 

\begin{figure}[!h]
    \centering
    \begin{subfigure}[c]{0.32\textwidth}
    \centering
    \includegraphics[keepaspectratio, height = 3cm]{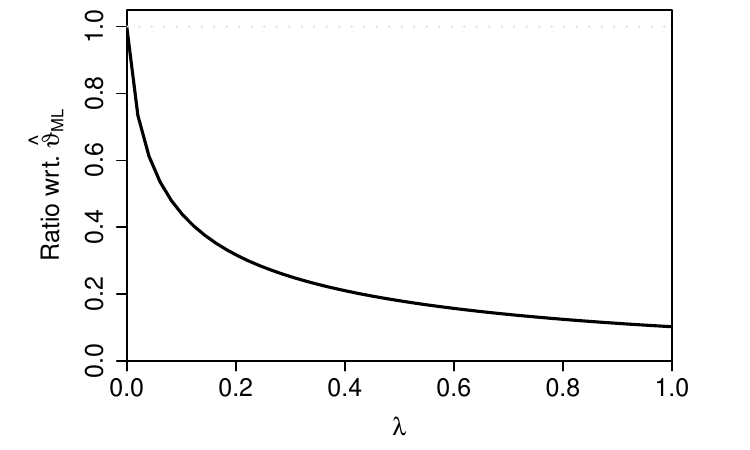}
    \caption{}
    \end{subfigure}
    \begin{subfigure}[c]{0.32\textwidth}
    \centering
    \includegraphics[keepaspectratio, height = 3cm]{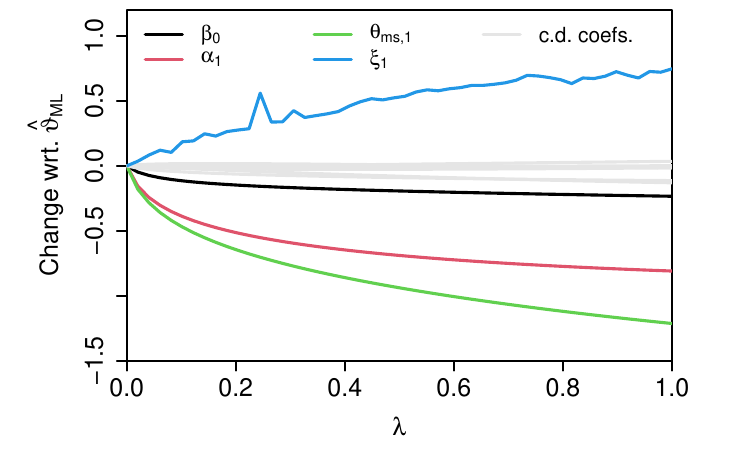}
    \caption{} 
    \end{subfigure} 
        \begin{subfigure}[c]{0.32\textwidth}
    \centering
    \includegraphics[keepaspectratio, height = 3cm]{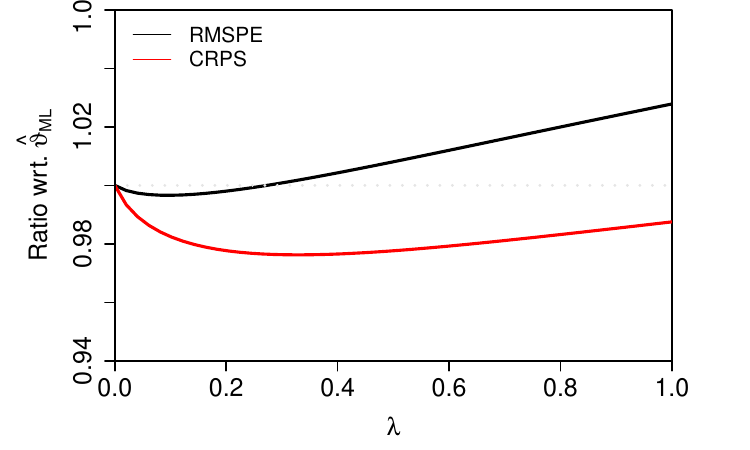}
    \caption{} 
    \end{subfigure} 
    \caption{Summary metrics under different penalization values. The condition number is presented in Panel~(a), relative change of $\hat{\bvartheta}_{pen}$ in Panel~(b), and prediction metrics in Panel~s(c). The c.d. coefs. label in Panel~(b) relates to those coefficients of the covariance function related to covariate-driven effects.}
    \label{fig:regularization_parameter}
\end{figure}

The regularization parameter $\lambda_r$ can be selected via grid search by optimizing an out-of-sample predictive criterion (e.g. RMSPE or CRPS). The goal is to balance the conditioning of the covariance matrix against the induced bias. Figure~\ref{fig:regularization_parameter} illustrates this trade‐off, where we model a nonstationary Gaussian process in the spatial trend, variance, and scale, shaped with sinusoidal covariates in each axis. As $\lambda_r$ increases, Panel~(a) shows a rapid decline in the covariance matrix condition number. Panel~(b) shows shrinkage primarily of the global scale and standard deviation parameters, with a compensatory increase in $\nu_0$ under weak penalization. 
%
In Panel (c), we observe that prediction accuracy (RMSPE and CRPS) remains within 3\% of the unpenalized model, occasionally improving due to reduced overfitting. In our bounded, single-realization setting, the likelihood surface typically exhibits a ridge over the $\theta_{ms,0}$ and $\alpha_0$ plane, leading to many (log)scale-(log)variance combinations yielding virtually the same covariance representation. As we increase $\lambda_r$ the optimal configuration for $\theta_{ms,0}$ and $\alpha_0$ drift along the weakly identifiable ridge between the (log)scale and (log)variance, leading to the overall fit to remain similar, so the covariate-driven coefficients, and the overall fit, remains similar as the non-penalized scenario. At the same time, applying a penalty to $\sqrt{\nu_0}$ prevents the smoothness $\nu_0$ from growing to compensate for the reduced scale, avoiding further numerical issues.

A key advantage of the presented covariance function is its parametric structure, which, following along the extensive theory in automatic model selection for parametric models, enables automatic variable selection in both the spatial mean and in each of the different models defining the nonstationary covariance function. As opposed to existing spatial modeling approaches, which rarely perform joint selection on mean and covariance terms, we consider an approach with separate Lasso penalties for the covariate effects in the mean and in each component of the nonstationary covariance function. To prevent numerical issues due to the non-differentiability of the $L_1$ penalty under gradient-based optimization such as \verb|L-BFGS-B|, we replace the absolute-value penalty with a smooth differentiable approximation. In particular, we employ the smooth $L_1$ function introduced in \citep{schmidt2007fast}
\begin{equation*}
    p(x;\kappa) = \kappa^{-1} [\log(1+ \exp(\kappa x)) + \log(1 + \exp(-\kappa x))],
\end{equation*}
which approximates the absolute value function for large $\kappa$ (i.e. $\kappa = 1\text{e}6$). This strategy is analogous to other smoothing approaches for $\text{L}_1$-regularization that enable efficient quasi-Newton optimization. By using a large $\kappa$ smooth surrogate, we retain the sparsity-inducing effect of the Lasso yet can safely apply standard gradient-based solvers without convergence problems.

Building on this framework, we propose a two-stage penalized likelihood procedure for model selection and estimation. In the first stage, we obtain a penalized maximum likelihood estimate $\hat{\bvartheta}_{s1}$ by maximizing 

\begin{equation*}
    l_{s1}(\bvartheta) = l_{pen}(\bvartheta) + n \lambda_{\mu} \sum_i p(\beta_i) + n \lambda_{\Sigma} \sum_j p(\vartheta_j),
\end{equation*}
yielding $\hat{\bvartheta}_{s1}$, and where $\lambda_{\mu}$ and $\lambda_{\Sigma}$ are separate Lasso tuning parameters for the covariate-driven spatial mean and covariance terms, respectively. 
Both Lasso hyperparameters can be chosen by model comparison criteria or cross-validation. This formulation uses hyperparameters to control how nonstationarity is distributed between the mean and covariance structures.

Solving the stage-one optimization yields an initial estimate $\hat{\bvartheta}_{s_1}$. Because the smooth $\text{L}_1$ penalty shrinks many coefficients towards zero without necessarily making them exactly zero, we apply a thresholding rule to determine the active set of selected parameters. In particular, we define the active support as $\text{Support}(\hat{\bvartheta}_{s_1})=\{,i:|\hat{\vartheta}_{s_1,i}|>\epsilon,\}$ for a small tolerance $\epsilon>0$. That is, any coefficient estimate whose magnitude is effectively zero (below $\epsilon$) is treated as absent from the model. This yields a reduced subset of covariate effects that are kept in the final model. In the second stage, we refit a reduced model containing only the parameters in this active set, while treating all other coefficients as zero. In order to do so, we maximize \eqref{eq:pen}, which helps mitigate the estimation bias induced by the Lasso penalization in the first stage. The final model is sparser, more interpretable nonstationary spatial model that retains only the covariate effects supported by the data, enabling automatic model selection without sacrificing predictive performance. 

\section{Illustration} \label{ch:illu}
In this section, we fit Gaussian process models to monthly precipitation data with covariance functions presented in \ref{sec:dense_model} and \ref{sec:sparse_model}. We assess their predictive performance based on held-out data against classical implementations, and against alternative models to evaluate the effect on prediction skills when considering a spatially-varying smoothness. The implementations rely heavily on the \verb|cocons| \texttt{R} package \citep{cocons}, which provides the statistical procedures to model and predict Gaussian processes with the presented class of covariance functions.

\subsection{Data}
We use data from the Copernicus Europe repository, which offers a wide range of down-scaled bioclimatic indicators at a $1 \times 1$ km resolution, derived from ERA5 and ERA5-Land reanalysis of a 40-year period (1979-2018) \citep{wouters2021downscaled}. 
Such datasets are extensively used in the biodiversity community for climate screening analyses and various downstream applications. Specifically, we work with data from Switzerland, contained within latitudes $45$\verb|°|$75'$ to $47$\verb|°|$93'$ and longitudes $6$\verb|°| to $10$\verb|°|$69'$, totalizing $N=69965$ observations.
We model the daily average precipitation for January for the period 1979-2018, shaping the nonstationarity in the spatial mean and covariance with bioclimatic indicators, as well as latitude, longitude, and elevation information collected from the \texttt{R} package \texttt{elevatr} \citep{elevatr}.
\begin{figure}
    \centering
    \includegraphics[keepaspectratio, height = 6cm]{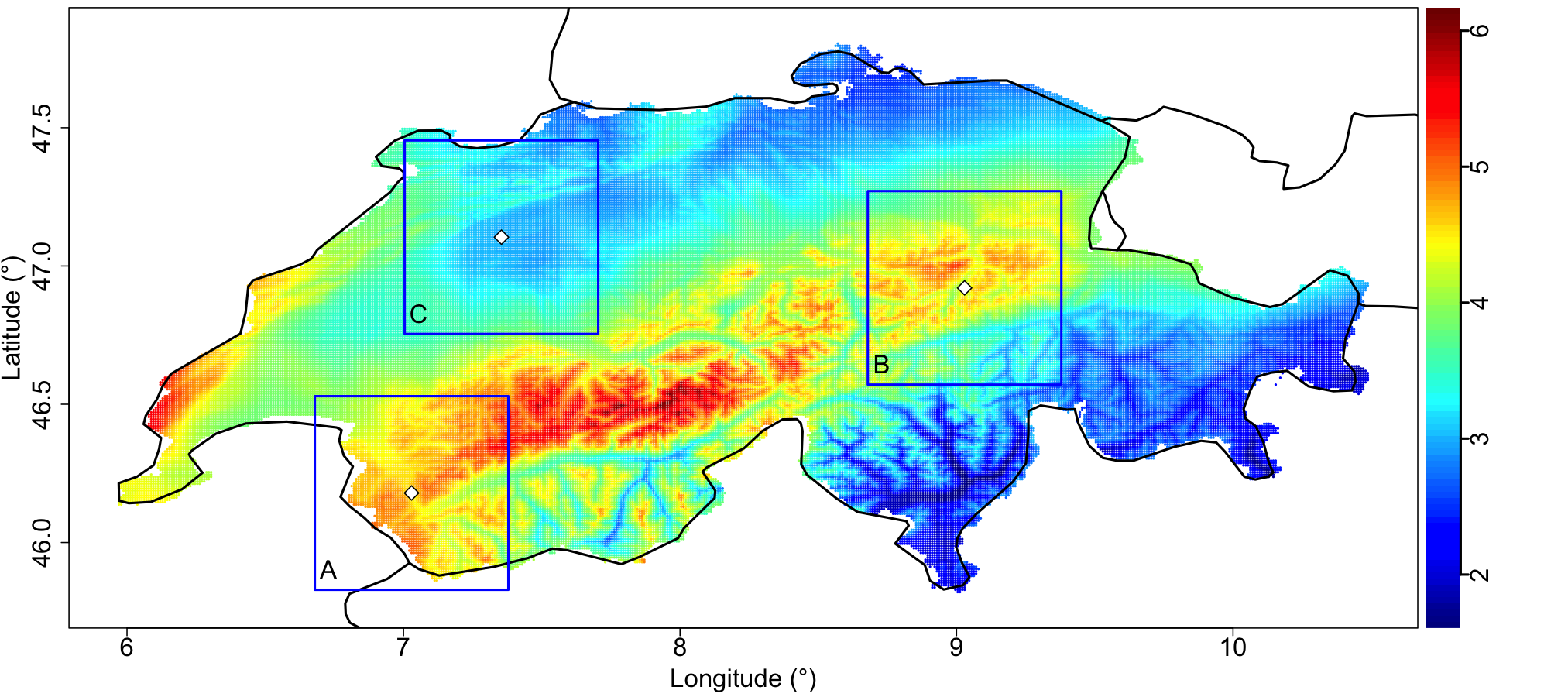}
    \caption{Average daily precipitation measured in millimeters for the month of January for the period 1979-2018 derived from ERA5. Regions within blue rectangles with respect to white diamonds at each center are inspected with a closer look.}
    \label{fig:prec}
\end{figure}
Instructions for accessing, downloading, and preprocessing the data are presented in Appendix~\ref{ap:access}.

\begin{table}[!h]
    \caption{Description of covariates}
    \begin{tabular}{lll}
    \hline
    Label & Variable & Description\\
    \hline
    \verb|prec| & Precipitation &  January daily average precipitation over the entire time period ($mm$) \\
    \verb|wind| & Wind & January daily average magnitude  over the entire time period of the two-dimensional \\ & &  horizontal air velocity near the surface over the entire time period ($ms^{-1}$) \\
    \verb|merwind| & Meridional wind speed & January daily average magnitude over the entire time period of the northward \\ & &  component of the two-dimensional horizontal air velocity near the surface ($ms^{-1}$)\\ 
    \verb|elev| & Elevation & Elevation ($mts$) \\
    \verb|BIO04| & Temperature seasonality & Standard deviation of the monthly mean temperature multiplied by 100 ($K$) \\
    \verb|BIO15| & Precipitation seasonality & Annual coefficient of variation of the monthly precipitation (-) \\
    \verb|cloud| & Cloud coverage & January daily average over the entire time period of the fraction of the grid for which \\ & &   the sky is covered with clouds. Clouds at  any height  above the surface are considered \\ & & (as a fraction) \\
    \end{tabular} \label{table:covariates}
\end{table}

\begin{figure}[!h]
    \centering
    \begin{subfigure}[c]{0.31\textwidth}
    \centering
    \includegraphics[keepaspectratio, height = 3.5cm]{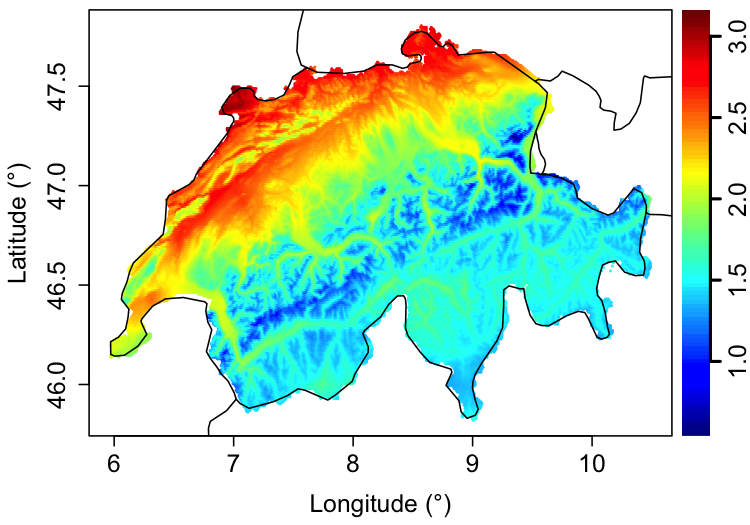}
    \caption{}
    \end{subfigure}
    \begin{subfigure}[c]{0.31\textwidth}
        \centering
        \includegraphics[keepaspectratio, height = 3.5cm]{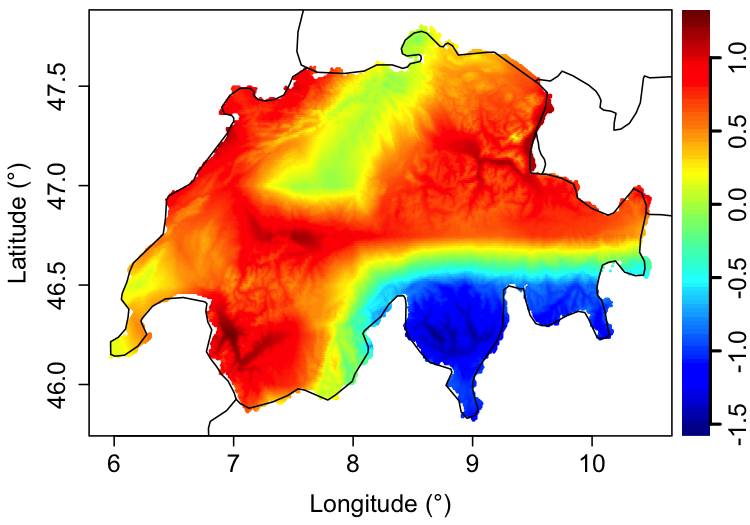}
    \caption{}
    \end{subfigure} 
    \begin{subfigure}[c]{0.31\textwidth}
        \centering
        \includegraphics[keepaspectratio, height = 3.5cm]{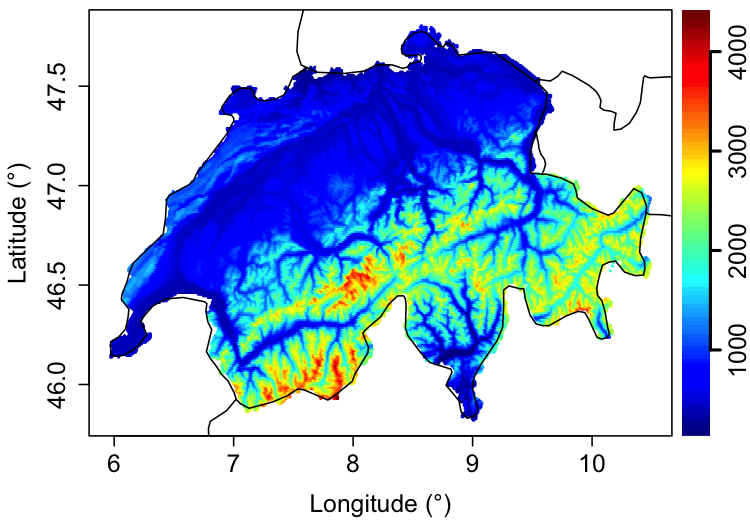}
    \caption{}
    \end{subfigure} \\ 
    \centering
    \begin{subfigure}[c]{0.31\textwidth}
    \centering
    \includegraphics[keepaspectratio, height = 3.5cm]{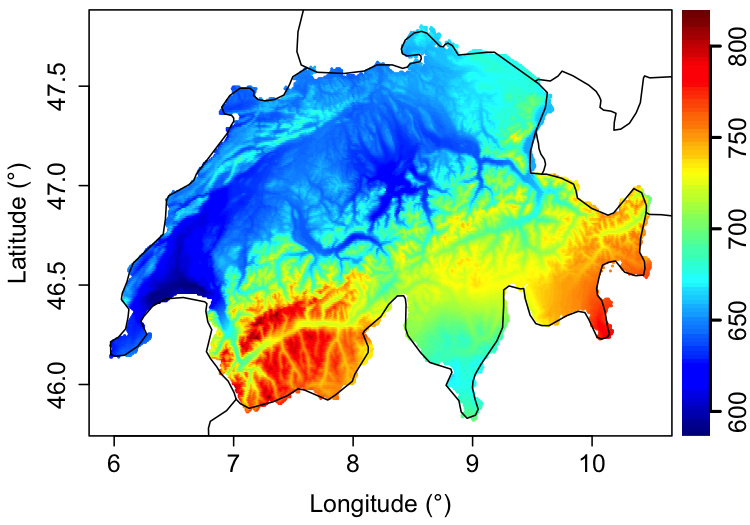}
    \caption{}
    \end{subfigure}
    \begin{subfigure}[c]{0.31\textwidth}
        \centering
        \includegraphics[keepaspectratio, height = 3.5cm]{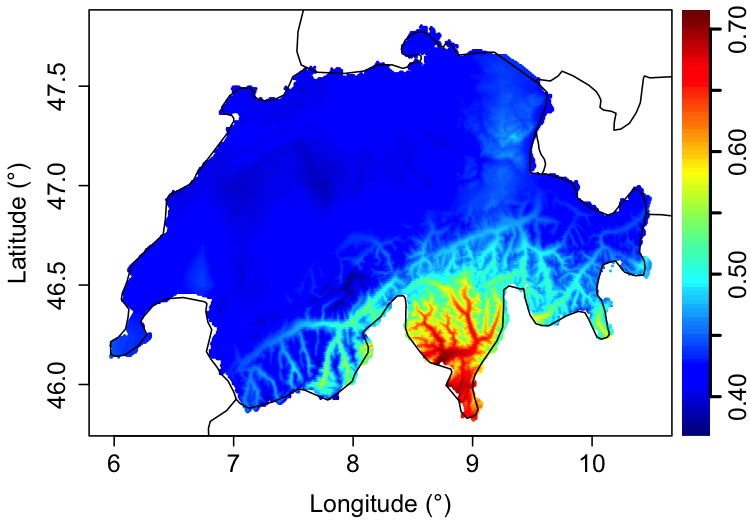}
    \caption{}
    \end{subfigure} 
    \begin{subfigure}[c]{0.31\textwidth}
        \centering
        \includegraphics[keepaspectratio, height = 3.5cm]{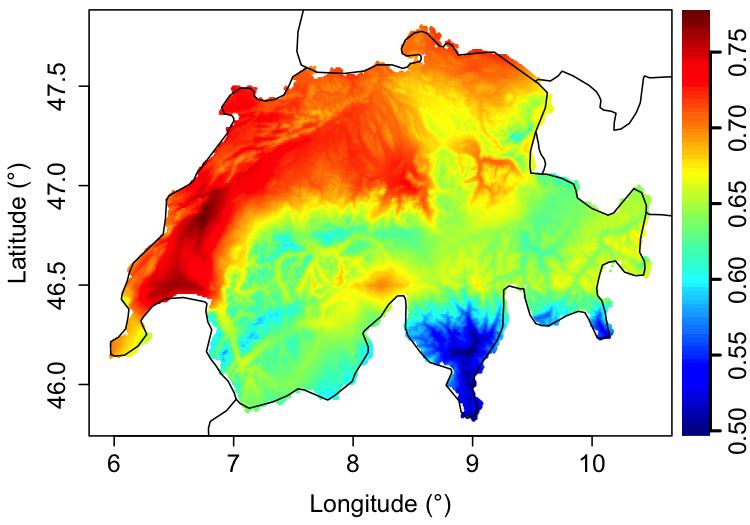}
    \caption{}
    \end{subfigure} 
    \caption{Down-scaled bioclimatic indicators: wind (a), meridional wind speed (b),  elevation (c), temperature seasonality (d),  precipitation seasonality (e), and cloud coverage (f). The heterogeneous covariates across Switzerland promote nonstationarity in both spatial trend and spatial structure.}
    \label{fig:bioclim_covariates}
\end{figure}

In Table~\ref{table:covariates}, Figure~\ref{fig:prec}, and Figure~\ref{fig:bioclim_covariates} we present an overview of the available covariates. Switzerland is characterized by its geographically diverse terrain, encompassing a mixture of wide, flat, low-altitude regions in the North with mountainous regions in the center and South, including the Alps and Jura Mountains, with narrow valleys dissecting these areas. When accounting for bioclimatic aspects such as cloud cover and wind patterns, distinct spatial structures with highly heterogeneous characteristics emerge. These factors greatly influence the spatial distribution of precipitation fields, requiring more flexible models.

\subsection{Framework} \label{sec:framework}
To analyze Gaussian process models with covariate-based covariance functions under both full (dense) and tapered settings, we consider two scenarios. In the Dense scenario, we subsample the training data to $n=500$ observations (taking every $k$-th observation) and use the covariance function described in Section 3.1. In the Sparse scenario, we use $n=10,000$ observations (again taking every $k$-th point) and apply a covariance tapering approach in Section~\ref{sec:sparse_model} to handle the larger dataset. To evaluate predictive performance, we utilize a hold-out set of $18781$ locations representing a mix of random points, linear stripes of varying widths, and small clusters. This diverse hold-out set ensures a thorough assessment of each covariance function across a range of spatial configurations. We allocate 30\% of these hold-out locations for tuning the three hyperparameters of the model (via a $3\times3\times3$ grid search minimizing the CRPS) and reserve the remaining 70\% for evaluating final predictive performance.

All candidate covariates are included as predictors in the spatial mean model as well as in each source of nonstationarity in the covariance structure. We standardize each covariate prior to modeling. For covariates included in the covariance function that take non-negative values, we first apply a logarithmic transformation to linearize their effects and then perform standardization.

We compare the performance of several Gaussian process models in this framework. The nonstationary models include M-NS (for the Dense scenario) and M-NS-T (for the Sparse scenario), each paired with a stationary counterpart (M-STAT and M-STAT-T, respectively).
A table summarizing the model structures and hyperparameters is provided in Appendix~\ref{ap:model_tables}.

\subsection{Evaluation criteria} \label{sec:criteria}
To assess the performance of the models on the hold-out dataset, we consider several criteria, including the Continuous Rank Probability Score (CRPS) and Log-Score \citep{gneiting2007strictly}, Root Mean Square Prediction Error (RMSPE), the Kolmogorov-Smirnov test statistics with respect to a standardized Gaussian distribution ($D_n$), as well as the empirical coverage probability of prediction intervals for a nominal level of $0.95$ (CPI).

We also report the number of parameters of the model, penalized log-likelihood values, and the computational time required to run the numerical optimizer \verb|L-BFGS-B|.
Instead of evaluating these criteria on the full hold-out sample, we split it into $100$ different sets of samples defined by a k-means algorithm creating heterogeneous scenarios to assess prediction capabilities of the models over a wide range of scenarios and spatial locations with heterogeneous characteristics. 
By calculating these criteria for each set, we account for the variability in prediction accuracy due to the selection of specific hold-out samples.

While RMSPE summarizes model quality in terms of bias, both the CRPS and Log-Score incorporate information about the uncertainty of the prediction distribution, making it more informative for comparing models with different spatial structures.
%
The \textit{best} score is achieved when the held-out data align perfectly with their predictive distributions \citep{gneiting2007strictly}.
For Gaussian processes, the CRPS at prediction location $\bs^p_{\ell}$ is defined as
\begin{equation}\label{eq:crps}
\text{CRPS}(\bs^p_{\ell}) = \sigma_{\ell} \left[\frac{1}{\sqrt{\pi}} - 2 \mathcal{N}_{\texttt{pdf}}\Bigl(\frac{z_{\ell}^p - \mu_{\ell}}{\sigma_{\ell}}\Bigr) - \Bigl(\frac{z_{\ell}^p - \mu_{\ell}}{\sigma_{\ell}}\Bigr)\Bigl(2\mathcal{N}_{\texttt{cdf}}\Bigl(\frac{z_{\ell}^p - \mu_{\ell}}{\sigma_{\ell}}\Bigr) - 1\Bigr)\right],
\end{equation}
where $\mu_{\ell}$ and $\sigma_{\ell}$ are the mean and standard deviation at prediction location $\bs_{\ell}^p$, and where $\mathcal{N}_{\texttt{pdf}}$ and $\mathcal{N}_{\texttt{cdf}}$ are the standardized univariate Gaussian density and cumulative functions, respectively. 
The Log-Score, on the other hand, takes the form of
\begin{equation}\label{eq:logscore}
    \text{Log-Score}(\bs^p_{\ell}) = \log(\sqrt{2\pi}) + \Biggl(\frac{z_{\ell}^p - \mu_{\ell}}{\sqrt{2} \sigma_{\ell}}\Biggr)^2 + \log(\sigma_{\ell}).
    \end{equation}
We estimate the CRPS and the Log-Score using plug-in estimates for the mean and variance of the predictive distribution.
We report the mean across holdouts for the CRPS, its empirical $0.95$ quantile, the Log-Score, $D_n$, and CPI. 

\subsection{Results}


After applying the two-step procedure and selecting hyperparameters to minimize the CRPS on a predefined grid, the dense nonstationary model M-NS retained only 20 of its original 49 parameters (effectively dropping $29$ parameters, or about $60\%$). Similarly, the sparse tapered model (M-NS-T) retained $16$ of its $33$ parameters (i.e., dropping $17$ parameters, or roughly $52\%$). In both models, the covariate-driven smoothness effects were shrunk to zero, collapsing $\nu$ to a single global value. In contrast, the spatially-varying scale and marginal standard deviation functions were retained in both models, suggesting these two sources of nonstationarity most contribute to capturing nonstationarity.
In terms of overall smoothness, both M-NS and M-NS-T yield similar estimates (approximately $1.7$ and $2.06$, respectively). These values are substantially higher than the smoothness estimate from M-STAT, which is around $0.9$. The lower smoothness in M-STAT seems to be a way to account for inadequacies in model fit, as highlighted in \citep{paciorek2003nonstationary}.

A visual representation of the models is presented in Figure~\ref{fig:var_wrong}, whereas a summary of parameter estimates is shown in Appendix~\ref{ap:model_tables}. Both M-NS and M-NS-T assign the largest spatial scale values to the Ticino region in southern Switzerland, consistent with Ticino’s Mediterranean-influenced microclimate and the broad, coherent rainfall events observed there. On the other hand, the central Alps exhibit much smaller scales, reflecting rather more localized mean precipitation behavior.

\begin{figure}[!h]
    \centering
    \begin{subfigure}[c]{0.31\textwidth}
    \centering
    \includegraphics[keepaspectratio, height = 3.5cm]{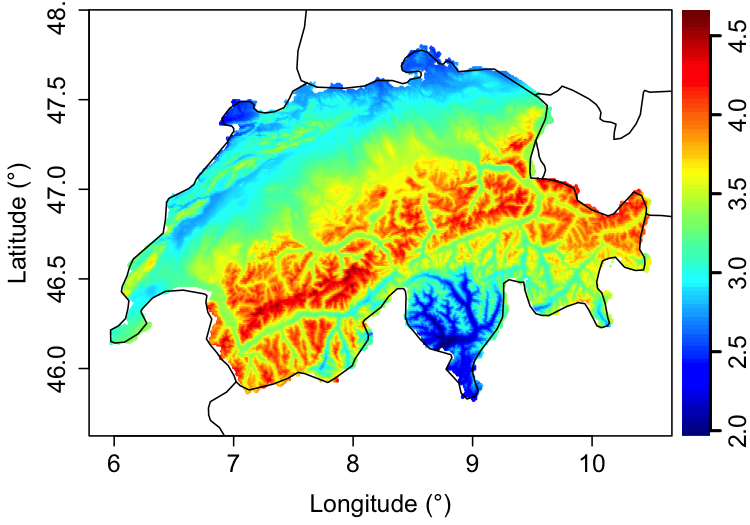}
    \caption{}
    \end{subfigure}
    \begin{subfigure}[c]{0.31\textwidth}
    \centering
    \includegraphics[keepaspectratio, height = 3.5cm]{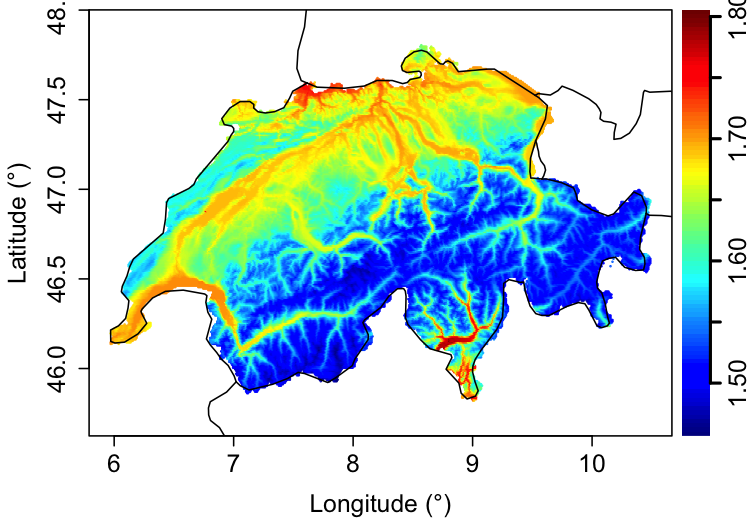}
    \caption{}
    \end{subfigure}
    \begin{subfigure}[c]{0.31\textwidth}
    \centering
    \includegraphics[keepaspectratio, height = 3.5cm]{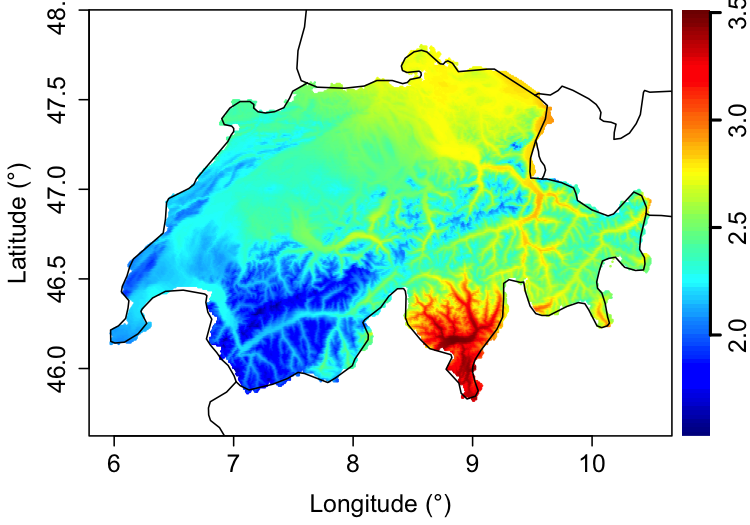}
    \caption{}
    \end{subfigure} \\ 
    \centering
    \begin{subfigure}[c]{0.31\textwidth}
    \centering
    \includegraphics[keepaspectratio, height = 3.5cm]{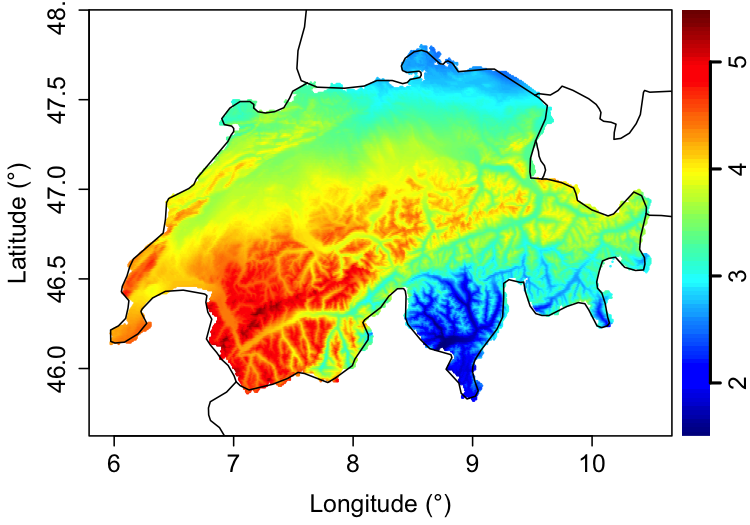}
    \caption{}
    \end{subfigure} 
    \begin{subfigure}[c]{0.31\textwidth}
    \centering
    \includegraphics[keepaspectratio, height = 3.5cm]{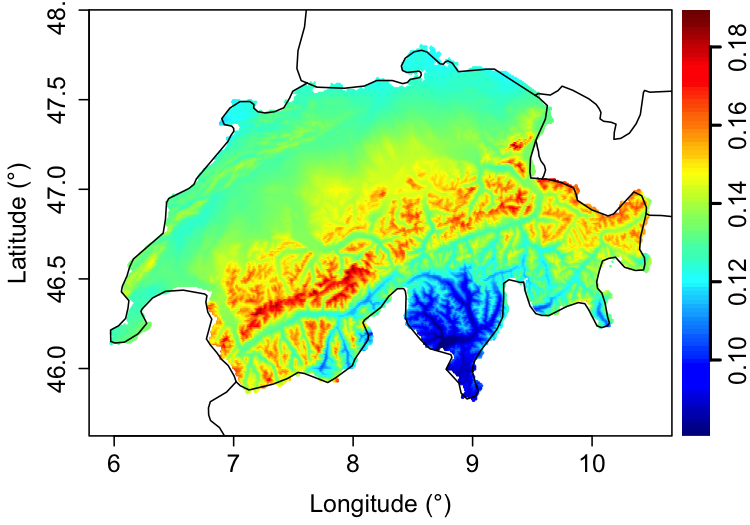}
    \caption{}
    \end{subfigure}
    \begin{subfigure}[c]{0.31\textwidth}
    \centering
    \includegraphics[keepaspectratio, height = 3.5cm]{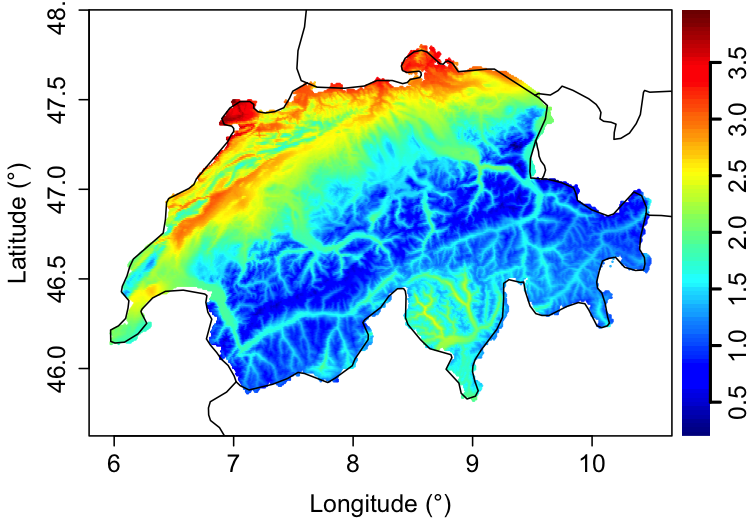}
    \caption{}
    \end{subfigure} 
    \caption{Spatial mean and nonstationary‐structure surfaces for the M-NS (first row) and M-NS-T (second row) models. First column shows the spatial mean over the full dataset. Second column shows the spatial surfaces of the marginal standard deviation and approximate local effective scale.}
    \label{fig:var_wrong}
\end{figure}

\begin{figure}[!h]
    \centering
    \begin{subfigure}[c]{0.33\textwidth}
    \centering
    \includegraphics[keepaspectratio, height = 3.5cm]{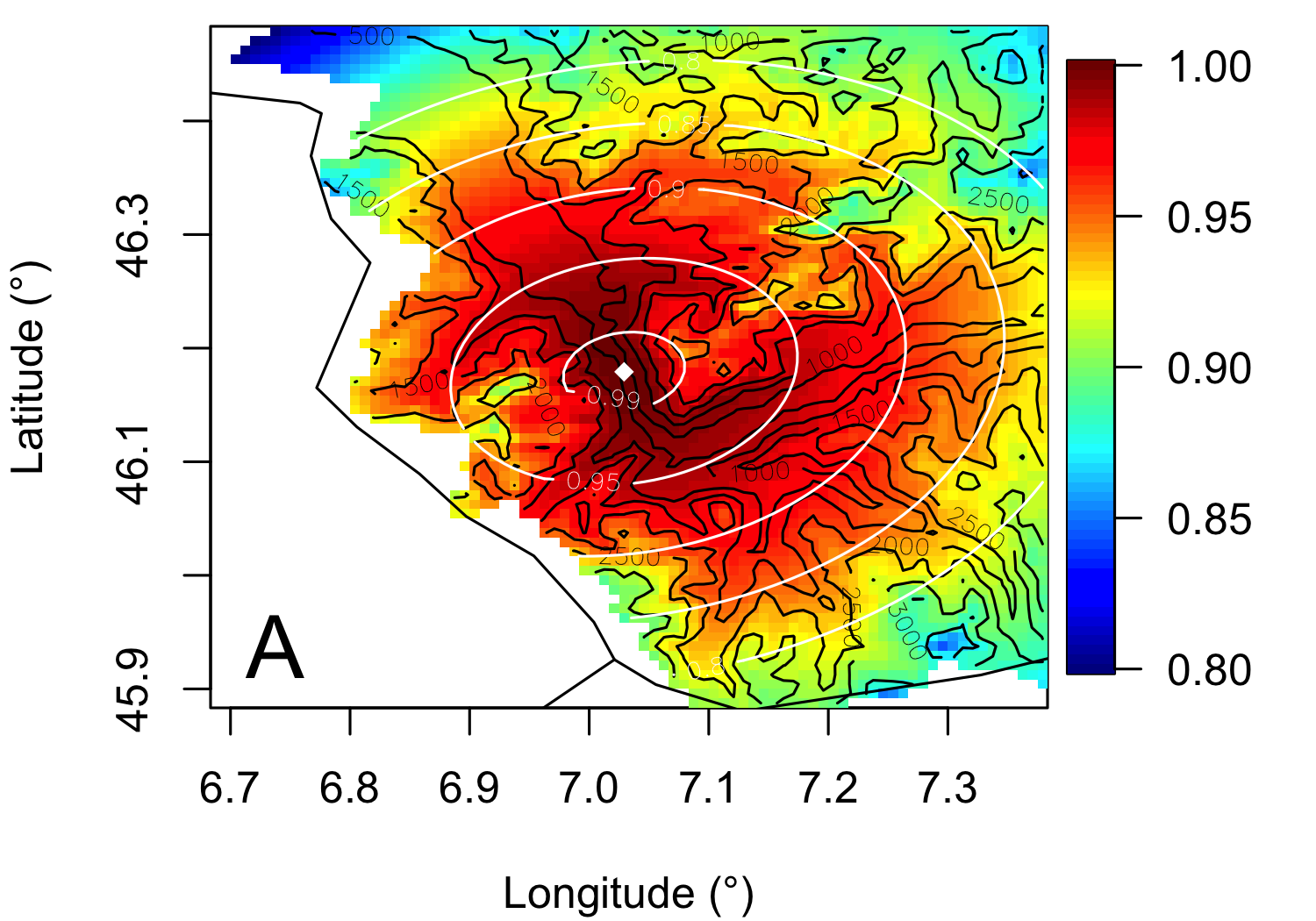}
    \caption{}
    \end{subfigure}
    \begin{subfigure}[c]{0.33\textwidth}
    \centering
    \includegraphics[keepaspectratio, height = 3.5cm]{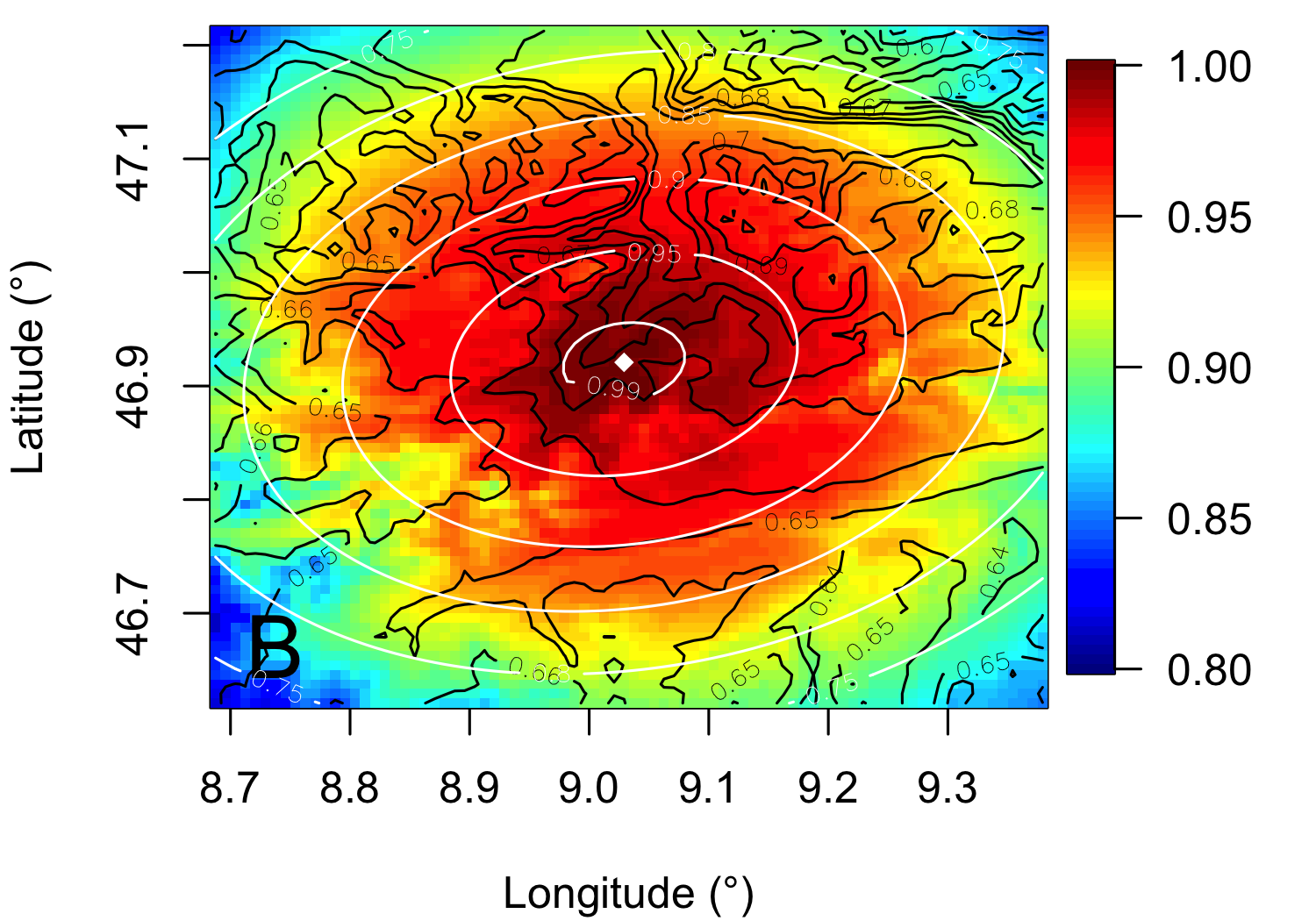}
    \caption{}
    \end{subfigure} 
    \begin{subfigure}[c]{0.33\textwidth}
    \centering
    \includegraphics[keepaspectratio, height = 3.5cm]{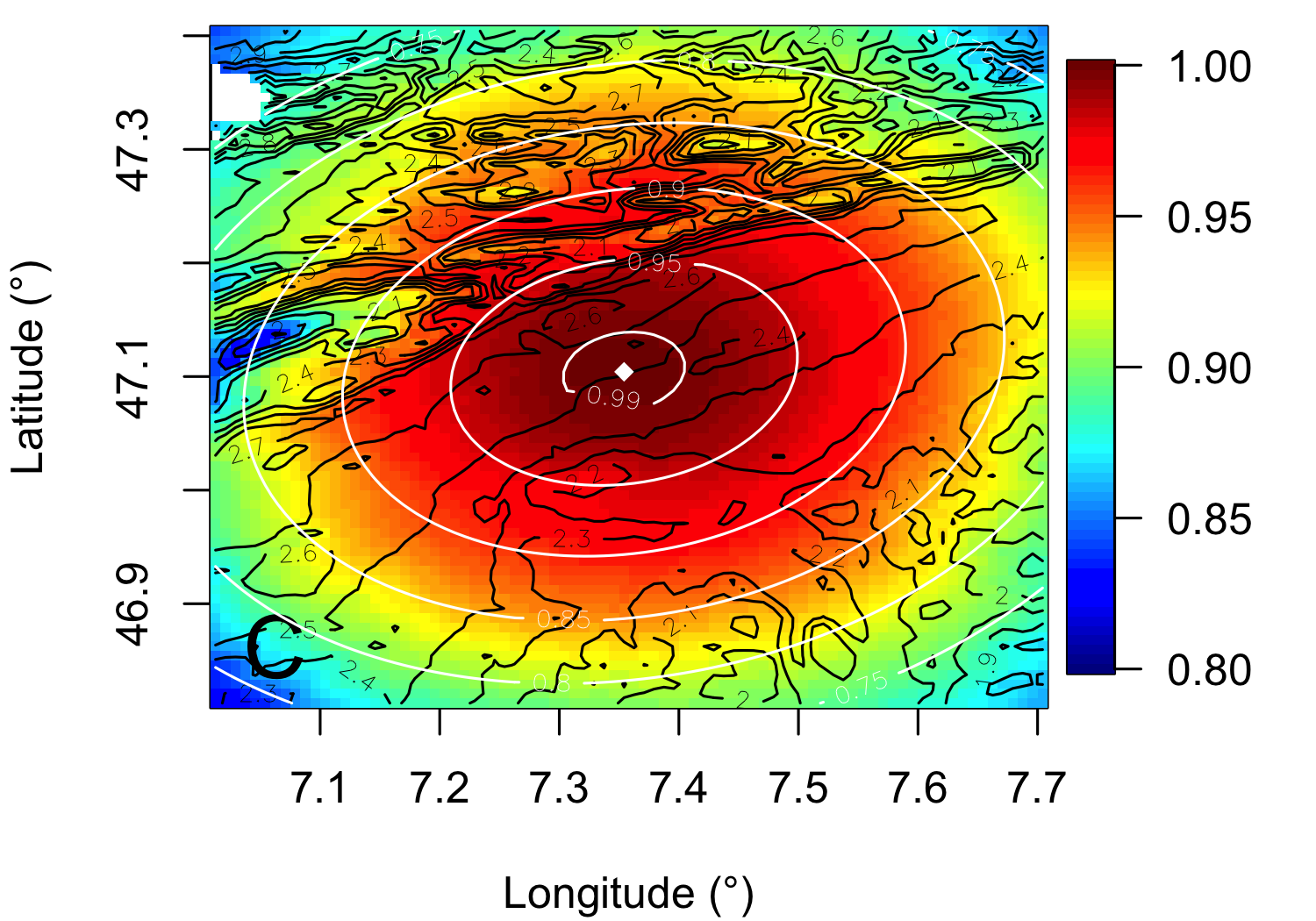}
    \caption{}
    \end{subfigure} \\ 
    \centering
    \begin{subfigure}[c]{0.33\textwidth}
    \centering
    \includegraphics[keepaspectratio, height = 3.5cm]{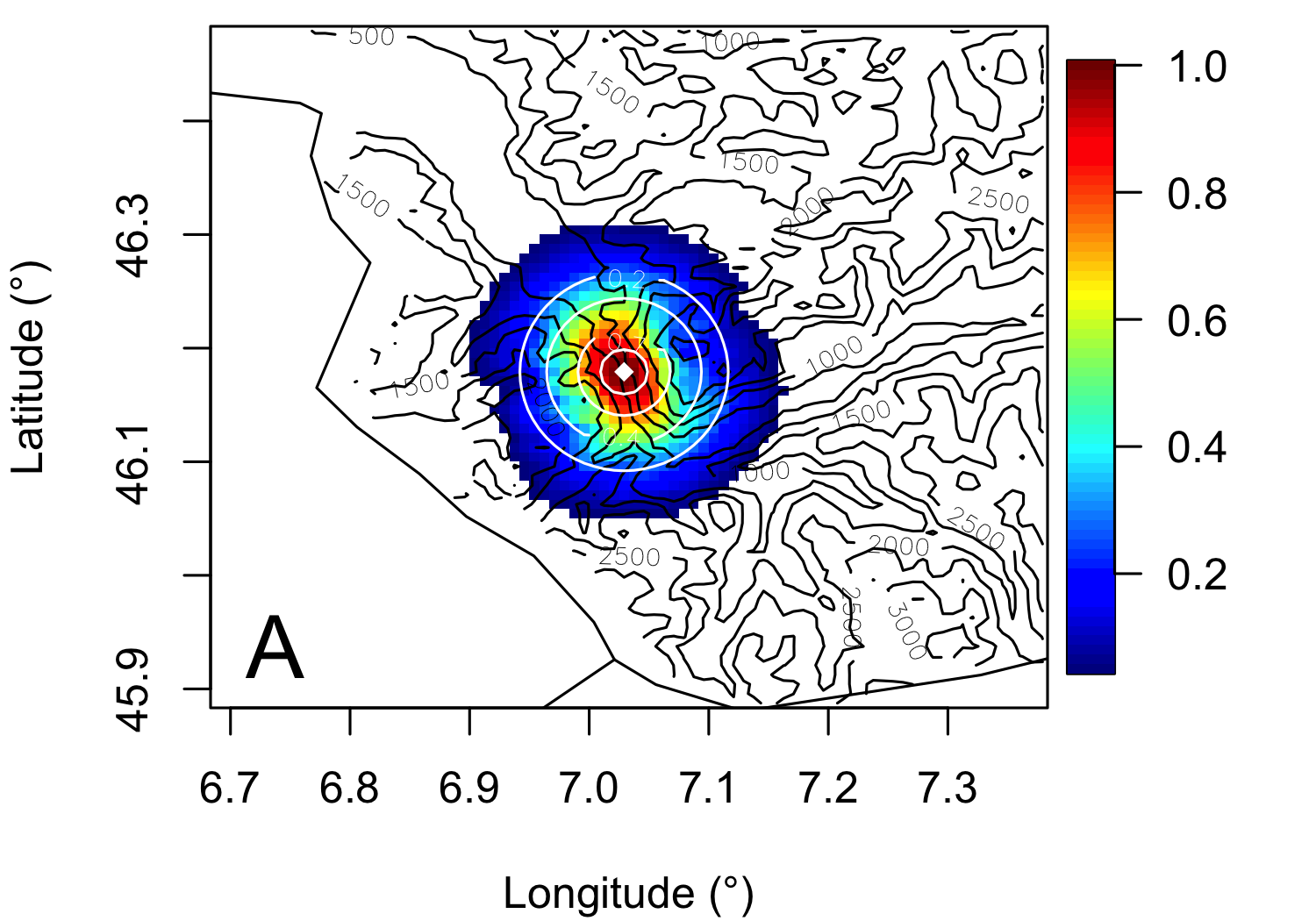}
    \caption{}
    \end{subfigure}
    \begin{subfigure}[c]{0.33\textwidth}
    \centering
    \includegraphics[keepaspectratio, height = 3.5cm]{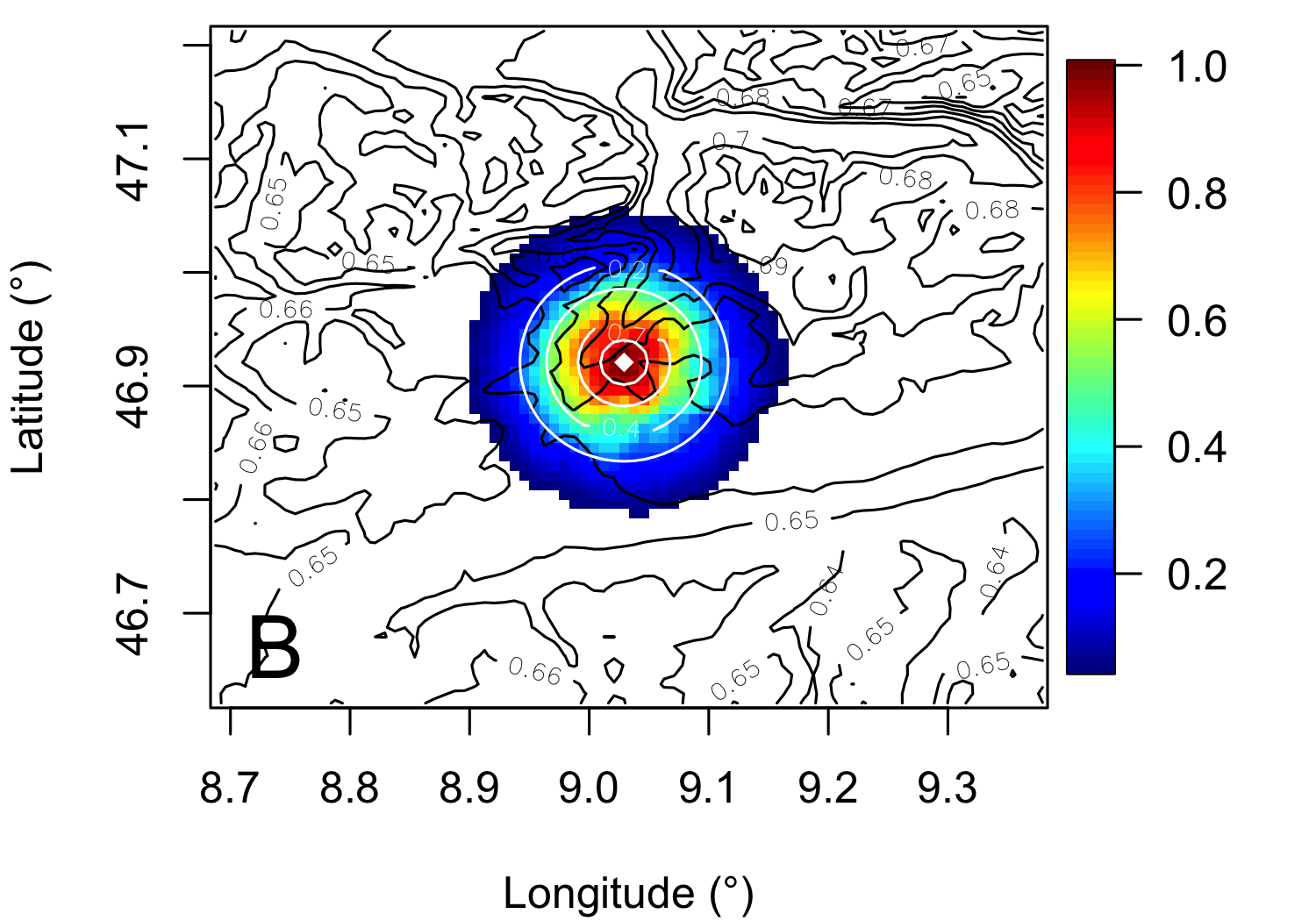}
    \caption{}
    \end{subfigure} 
    \begin{subfigure}[c]{0.33\textwidth}
        \centering
        \includegraphics[keepaspectratio, height = 3.5cm]{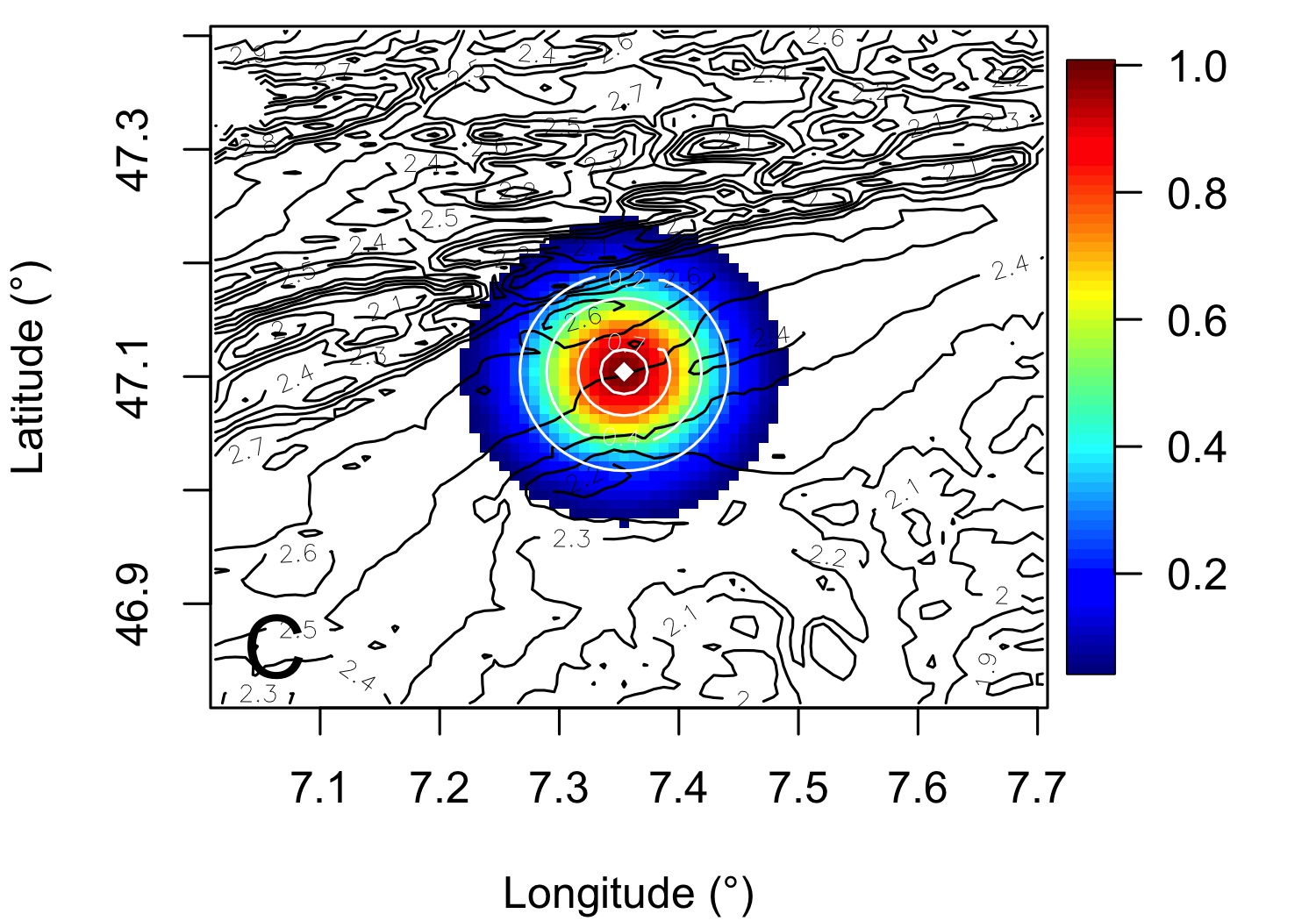}
        \caption{}
        \end{subfigure} 
    \caption{Correlation plots for M-NS (first row) and M-NS-T (second row) at three different regions, with correlation isolines (in white) of M-STAT (first row), and tapered correlation model M-STAT-T (second row). In black, countours with respect to different covariates. 
    Panels (a) and (d) relate to a location in a narrow valley characterized by large wind, surrounded by mountainous territory, with contours describing elevation. In (b) and (e), the contours describe cloud coverage, while in (c) and (f), wind.}
    \label{fig:correlations}
\end{figure}

In Figure~\ref{fig:correlations}, we compare the correlation structures for M-NS and M-NS-T across three distinct regions. Each column relates to the different boxes in Figure~\ref{fig:prec}.
The added flexibility of both models allows the spatial structure to adjust locally to the characteristics of the spatial locations, without forcing nonstationarity in regions where a simple stationary structure suffices.
Focusing on the first row (M-NS), Panel~(a) considers a location (white mark) in a narrow valley with steep surrounding topography. Here, the correlation structure concerning the white mark location reduces drastically to the East-West due to the steep increase in elevation while keeping a high correlation within the valley locations aligned in the North-South direction. This behavior is in stark contrast to the fixed global anisotropy of the stationary model M-STAT, which cannot account in the covariance structure for the steep boundaries of the valley mantaining relatively high correlations in all directions.
By contrast, Panels~(b) and (c) show that at their respective marked locations, the M-NS correlation structures are much closer to those of M-STAT. Only smooth, minor adjustments appear in these cases, and in Panel~(c) the M-NS correlation map almost perfectly agrees with its stationary counterpart, indicating that the local geographic and climatic context there does not demand a strong deviation from a simple, global anisotropic structure.
A similar pattern is observed in the second row of Figure~\ref{fig:correlations} for the taper-based models. In M-NS-T, the correlation function is bounded by the fixed taper scale, which forces the correlation to drop to zero once the distance-based quantity $Q_{ij}$ approaches the threshold $\delta=0.23$. Despite this imposed cutoff, the Panels~(d–f) for M-NS-T exhibit the same kind of localized shape adjustments as in M-NS, and the intensity of these adjustments is preserved across the three regions. In other words, even with the taper constraint, the nonstationary model adapts its correlation structure to local features in much the same way, yielding nearly the same pattern of anisotropy adjustments in each region as we saw in the untapered case. In both the tapered and untapered formulations, the nonstationary correlation structure adapts locally to terrain and climate features without unnecessary complexity elsewhere.

\begin{table}[ht]
    \centering
    \caption{Summary of performance metrics. Standard errors between holdouts are shown in parentheses. Bold text indicates the best metric achieved in each scenario. Time is presented in minutes.}
    \begin{tabular}{c|cc|cc}
    \multirow{2}{*}{Metric} 
      & \multicolumn{2}{c|}{Dense \textit{(n=500)}} 
      & \multicolumn{2}{c}{Sparse \textit{(n=10000)}} \\
    \cline{2-5}
      & M-STAT & M-NS 
      & M-STAT-T & M-NS-T \\
    \hline
    RMSPE              & 0.069 \textit{(0.032)} & \textbf{0.039} \textit{(0.028)} 
                       & 0.040 \textit{(0.018)} & \textbf{0.013} \textit{(0.010)} \\
    CRPS               & 0.045 \textit{(0.034)} & \textbf{0.022} \textit{(0.018)} 
                       & 0.024 \textit{(0.021)} & \textbf{0.006} \textit{(0.005)} \\
    $q_{0.95}(\mathrm{CRPS})$ 
                       & 0.095 \textit{(0.056)} & \textbf{0.052} \textit{(0.037)} 
                       & 0.063 \textit{(0.047)} & \textbf{0.019} \textit{(0.013)} \\
    Log-Score          & -1.087 \textit{(0.106)} & \textbf{-2.105} \textit{(0.096)} 
                       & -2.161 \textit{(0.623)} & \textbf{-3.534} \textit{(0.077)} \\
    $D_n$              & \textbf{0.298} \textit{(0.152)} & 0.299 \textit{(0.167)} 
                       & 0.229 \textit{(0.085)} & \textbf{0.228} \textit{(0.089)} \\
    CPI                & \textbf{0.923} \textit{(0.108)} & 0.900 \textit{(0.114)} 
                       & \textbf{0.938} \textit{(0.077)} & 0.972 \textit{(0.043)} \\
    \hline
    $\dim(\bvartheta)$           & 14  & 20  
                                  & 12  & 16   \\
    $l_{pen}(\hat{\bvartheta}_{pen})$ 
                                  & -971  & -1564  
                                  & -36043 & -61995 \\
    time                       & 0.62 & 1.44  
                                & 8.00 & 12.68 \\
    \end{tabular}
    \label{tab:pred_summary}
\end{table}

A summary of prediction metrics as well as computational, loglikelihood, time and number of parameters is presented in Table~\ref{tab:pred_summary}.
The training and test values, as well as the predictive mean and standard for each of the considered models is presented in Appendix~\ref{ap:illu_figures}.
The nonstationary models (M-NS and M-NS-T) show a clear advantage in accuracy over their stationary counterparts across almost all scoring metrics. 
M-NS and M-NS-T have lower RMSPE, CRPS, and log-predictive scores than M-STAT and M-STAT-T, respectively, indicating that their predictions are more precise, providing more adequate uncertainty quantifications, and where the worse 5\% of the CRPS distributions are almost more than half as small as their classical counterparts, meaning more robust uncertainty quantification under a heterogeneous number of hold-out sets. 
The improvements for M-NS and M-NS-T in these metrics are substantial, with M-NS showing an approximately 50\% reduction compared to M-STAT and an impressive 75\% reduction in CRPS for M-NS-T compared to M-STAT-T. 
When comparing across scenarios, it is notable that M-NS nearly matches the predictive performance of M-STAT-T, even though M-STAT-T was trained on twenty times more observations (leading to closer training points for each prediction site).
These results indicate a clear advantage of allowing nonstationary covariance structures for spatial interpolation and uncertainty quantification. 
The added flexibility of the proposed nonstationary spatial model translates into sharper and more accurate predictive distributions on held-out data. Moreover, the Kolmogorov–Smirnov goodness-of-fit statistic ($D_n$) reveals comparable distributional fit between stationary and nonstationary models within each scenario, indicating that the more complex models do not degrade the overall distributional fit.

One trade-off observed is that nonstationary models can exhibit slightly worse coverage of the nominal prediction intervals relative to the stationary models. For the dense scenario, M-NS empirical coverage lies considerably below the nominal value (at $90\%$), slightly lower than M-STAT at $92.3\%$. In the sparse scenario, this trend reverses, M-NS-T achieves about $97.2\%$ coverage versus $93.8\%$ for M-STAT-T, slightly over-covering the nominal $95\%$. This pattern might be explained by the fact that nonstationary models have been selected by hyperparameter tuning prioritizing the CRPS, favoring sharper predictive distributions rather than coverage. 

Regarding computational time, nonstationary models incur only a modest overhead in both scenarios and after the two-stage penalization, they run in the same order of magnitude as their stationary counterparts, offering an excellent trade-off between computation and predictive improvements.

\section{Discussion} \label{ch:sec5}
\newcommand{\ceil}[1]{\left\lceil #1 \right\rceil}
In this article, we presented a class of parametric modular covariate-based covariance functions that, by leveraging observable spatial covariates, is able to represent nonstationary spatial processes capable of achieving flexibility while offering an economical parameterization, keen to computational efficiency. 
It allows the representation of up to five sources of nonstationarity, including marginal standard deviation, a three-component local geometric anisotropy, and smoothness, all introduced modularly. 
We introduced a tailored regularization strategy that promotes well-behaved covariance estimates without sacrificing predictive power, as well as a two-stage estimation approach for automatic model selection. 
The Mat\'ern covariance function is nested into the nonstationary model presented in \eqref{eq:gen_rns_ours}, and can be adapted for large datasets enhancing flexibility and computational efficiency for a wide spectrum of sample sizes. Moreover, it offers a plethora of numerical and visual tools to explore the contributions of each source of nonstationarity.

On a challenging reanalysis data example, the nonstationary models produce more meaningful and sensible results than those from a stationary model. 
The results over seventy heterogeneous hold-outs sets revealed that the nonstationary models delivered better prediction distributions when compared to classical stationary implementations, at only a minor increase in computational cost. Moreover, when considering the comparable computational times against their stationary counterparts, nonstationary models offer an excellent trade-off between flexibility and efficiency for moderate sample sizes. There are, however, alternative strategies for scaling Gaussian process models to very large datasets.
Alternative approaches can be to use full-scale approximations (FSAs) combining predictive process methods and covariance tapering \citep{gyger2024iterative}, to work with multi-resolution approximations (M-RA), allowing local nonstationary covariance functions \citep{huang2021nonstationary}, or to consider highly-scalable maximum weighted composite likelihood based on pairs (WCLP) with symmetric weight function based on nearest neighbors \citep{caamano2024nearest}.

There was little justification to incorporate an extra spatially-varying smoothness function on top of other spatially-varying models when the goal is to perform predictions. 
After our two-stage model selection procedure, both nonstationary candidate models retained only a single global smoothness parameter, suggesting that allowing smoothness to vary spatially did not improve predictive performance.
One factor explaining this limited improvement is related to the dataset employed in the illustration, which (although openly accessible) might not be sensible to models representing spatially-varying smoothness since the resulting high-resolution grid of observations is, in fact, solutions of Delaunay linear interpolation from weather stations, and thus may not reflect genuine small-scale differences in the underlying smoothness of the process. Other data sets might be more suitable for benefiting from models with spatially-varying smoothness, such as in \citet{fang1998some}, where the smoothness of longitudinal variations in total column ozone in the Earth's atmosphere shows a clear dependence on latitude.

Our proposed nonstationary covariance structure offers a layer of interpretability, but this must be viewed in light of practical identifiability limitations. Structurally, the model is interpretable by construction: each covariate is linked to a specific aspect of the covariance (variance, anisotropy, smoothness, etc.), which in principle allows one to attribute changes in correlation structure to particular spatial features. However, in practice we are inferring many parameters from a single spatial field, which makes certain effects only weakly identifiable. Consequently, the estimation results can be sensitive to the choice of initial values. We expect this situation to improve if multiple independent realizations of the process are available or if the model is extended to a spatio-temporal context, where more information helps stabilize parameter estimation.

On the positive side, our penalization and model-selection approach tends to eliminate unwarranted complexity: unlike methods that impose nonstationarity regardless of data support, our procedure dropped any nonstationary components that were not needed to adequately describe the data. For example, in both of our nonstationary models the initially full covariate-driven smoothness model collapsed to a single global smoothness term, and all tilt parameters and the majority of scale‐and‐anisotropy covariate effects were driven to zero, leaving only a small subset of covariate effects that meaningfully improved fit. This data-driven parsimony partly addresses the question of where and to what extent nonstationarity is required in the covariance function, as noted by \cite{fuglstad2015does}.

The Taper approach strategy presented in Section~\ref{sec:sparse_model} greatly reduces computational burden while still allowing for capturing spatially‐varying covariance structure. We employ a single, isotropic taper scale as a practical convenience: once the sparse matrix pattern is constructed, it can be reused across all likelihood evaluations. However, this choice may be suboptimal in regions where the true correlation scale varies substantially. Spatially adaptive tapering \citet{bolin2016spatially}, allows the taper scale itself to vary with location, enabling more targeted sparsity and improving local approximation. In practice, adaptive tapering requires recomputing the permutation and symbolic factorization for each new taper pattern, incurring up to $O(n^{3/2})$ cost per iteration and eroding much of the computational advantage. We therefore leave spatially‐varying taper scales as an avenue for future work, trading off local fidelity against the benefits of a fixed, reusable sparse structure.

Finally, with regard to regularization and optimization, one could avoid our smooth-$L_1$ approximation and post-hoc thresholding by using an optimizer specifically designed for non-differentiable $L_1$ objectives. In particular, the Orthant-Wise Limited-memory Quasi-Newton (OWL-QN) algorithm \citep{andrew2007scalable} directly handles the $L_1$ penalty and can yield exact zero estimates, eliminating the need to soften the Lasso penalty. OWL-QN does require access to the gradient of the smooth part of the objective, but in our scenario, it might be obtainable via automatic differentiation \citep{baydin2018automatic}. Integrating OWL-QN with modern autograd tools is therefore a promising direction for providing a mathematically rigorous and computationally robust avenue for future methodological development.

\section{Appendix}

\subsection{Positive definiteness of the presented modular covariance function} \label{ap:proof}
In this appendix, we show that the nonstationary covariance function as in~\eqref{eq:c_ns}, with $\nu(\bs_i,\bs_j;\bxi) = \sqrt{\nu_i \nu_j}$, where $\nu_i = \nu(\bs_i;\bxi)$ follows \eqref{eq:spatially_var_smtns} is positive definite. 
The proof is a simple extension of \citep{anderes2011local}, which shows that the covariance function \eqref{eq:c_ns}, with spatially-varying smoothness $\nu(\bs_i,\bs_j; \xi) = (\nu_i + \nu_j)/2$ is positive definite.
We start by introducing the lemma by \citep{anderes2011local}, which is then used to show positive definiteness on the introduced covariance function \eqref{eq:c_ns} with $\nu(\bs_i,\bs_j;\bxi) = \sqrt{\nu_i \nu_j}$, where $\nu_i = \nu(\bs_i;\bxi)$ follows \eqref{eq:spatially_var_smtns}.
\begin{lemma}
    Let $\bSigma(\cdot;\bs): \mathbb{R}^p \to d\times d $ real positive-definite matrices, 
    $\sigma(\cdot;\bs): \mathbb{R}^p \to \mathbb{R}$, 
    and for each $\bs \in \mathbb{R}^p, g(\cdot;\bs) \in L^2(dH)$, being $H$ nonnegative and bounded on $[0,\infty)$. Then
    \begin{equation}\label{eq:the_cov}
        \Cov(\bs_i, \bs_j) = \sigma_i \sigma_j \frac{|\bSigma_i|^{1/4} |\bSigma_j|^{1/4}}{\bigl|\frac{\bSigma_i + \bSigma_j}{2}\bigr|^{1/2}} \int_{0}^{\infty} \exp(-Q_{ij} w) g(w;\bs_i) g(w;\bs_j) dH(w)
    \end{equation}
    is positive definite. By taking $dH(w) = w^{-1}\exp(-1/(4w)) dw$ with respect to Lebesgue measure, $g(w;\bs) = w^{-\nu(\bs)/2}$, by using a convolution argument from (\cite{paciorek2003nonstationary}, p.23), and by using (3.471.9) in \cite{gradshteyn2014table} it can be shown the resulting covariance function is positive definite and leads to \eqref{eq:c_ns} with $(\nu_i + \nu_j)/2$ (\cite{stein2005nonstationary}, p.3).
\end{lemma}

\noindent\textbf{Claim.} The covariance function \eqref{eq:c_ns} with $\nu(\bs_i,\bs_j;\xi) = \sqrt{\nu_i \nu_j}$, where $\nu_{\ell} = \nu(\bx_{\ell};\bxi)$, is defined as in \eqref{eq:spatially_var_smtns} being $\numin \leq \nu_{\ell} \leq \numax$ leading to
\begin{equation}
    \Cov_{R}(\bx_i, \bx_j) = \sigma_i \sigma_j \frac{|\bSigma_i|^{1/4} |\bSigma_j|^{1/4}}{\bigl|\frac{\bSigma_i + \bSigma_j}{2}\bigr|^{1/2}} \int_{0}^{\infty} \exp(-Q_{ij}w) h(w;\bx_i,\bx_j) dH(w),
\end{equation}
with $h(w;\bx_i,\bx_j) = w^{-\sqrt{\nu_i \nu_j}}$, is positive definite.

\begin{proof}
    We define the piecewise function
    \begin{equation*}
        m(w;\bs)  = \begin{cases}
            w^{-\frac{\numin}{2}}, & \text{if} w \leq 1 \\
            w^{-\frac{\numax}{2}}, & \text{otherwise}
        \end{cases},
    \end{equation*}
    which for each $\bs$, $m(\cdot;\bs) \in L^2(dH)$. Notice then that $h(w;\bs_i,\bs_j) \geq m(w;\bs_i) m(w;\bs_j), \forall w \in \mathbb{R}^+$. Then, using the convolution argument (\cite{paciorek2003nonstationary},p. 27),
    \begin{align*}
        \sum_{i=1}^n \sum_{j=1}^n c_i c_j \Cov_{R}(\bs_i,\bs_j) &= 
        \sum_{i=1}^n \sum_{j=1}^n c_i c_j  \sigma_i \sigma_j \frac{|\bSigma_i|^{1/4} |\bSigma_j|^{1/4}}{\bigl|\frac{\bSigma_i + \bSigma_j}{2}\bigr|^{1/2}} \int_{0}^{\infty} \exp(-Q_{ij}w) h(w;\bs_i,\bs_j) dH(w) \\
        &\geq  \sum_{i=1}^n \sum_{j=1}^n c_i c_j \sigma_i \sigma_j \frac{|\bSigma_i|^{1/4} |\bSigma_j|^{1/4}}{\bigl|\frac{\bSigma_i + \bSigma_j}{2}\bigr|^{1/2}}  \int_{0}^{\infty} \exp(-Q_{ij}w) m(w;\bs_i) m(w;\bs_j) dH(w) \\
        &= \Bigl(2\sqrt{\pi}\Bigr)^{d/2} \sum_{i=1}^n \sum_{j=1}^n c_i c_j \sigma_i \sigma_j |\bSigma_i|^{1/4} |\bSigma_j|^{1/4} \int_{0}^{\infty} \Biggl(\int_{\cD} K_{\bs_i}(\bu) K_{\bs_j}(\bu) d\bu\Biggr) m(w;\bs_i) m(w;\bs_j) dH(w) \\ 
        &= \Bigl(2\sqrt{\pi}\Bigr)^{d/2} \int_{0}^{\infty} \int_{\cD} \Bigl(\sum_{i=1}^n c_i \sigma_i |\bSigma_i|^{1/4} K_{\bs_i}(\bu) m(w;\bs_i)\Bigr)^2 d\bu dH(w) \geq 0 
    \end{align*}

which is nonnegative, proving that $\Cov_{R}(\bs_i,\bs_j)$ assign nonnegative values to all quadratics forms. Then, starting from \eqref{eq:the_cov} and by using (3.471.9) in \cite{gradshteyn2014table}, it can be shown the resulting positive definite covariance function leads to 
\begin{equation}
    \Cov_{R}(\bs_i, \bs_j) = \sigma_i \sigma_j \frac{|\bSigma_i|^{1/4} |\bSigma_j|^{1/4}}{\Bigl|\frac{\bSigma_i + \bSigma_j}{2}\Bigr|^{1/2}} \cM_{\sqrt{\nu_i \nu_j}}\Bigl(\sqrt{Q_{ij}}\Bigr)
\end{equation}
\end{proof}

\subsection{Illustration figures} \label{ap:illu_figures}

The following Figures provided added information to the illustration section.

\begin{figure}[!h]
    \centering
    \begin{subfigure}[c]{0.33\textwidth}
    \centering
    \includegraphics[keepaspectratio,height=3.5cm]{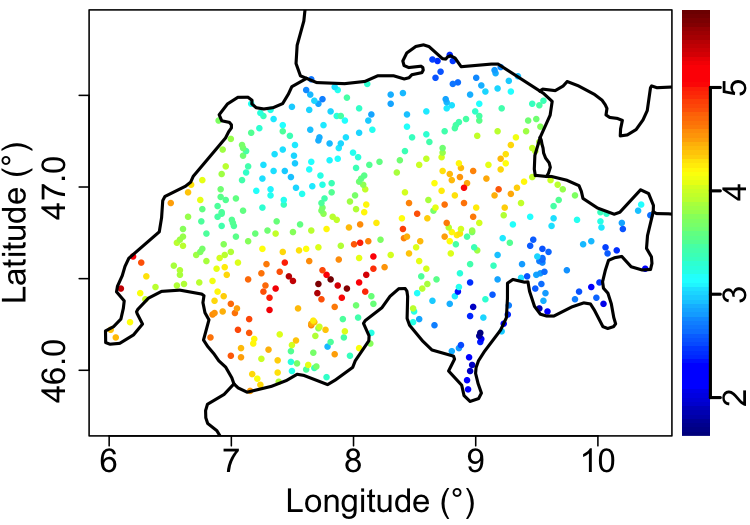}
    \caption{}
    \end{subfigure}
    \begin{subfigure}[c]{0.33\textwidth}
    \centering
    \includegraphics[keepaspectratio,height=3.5cm]{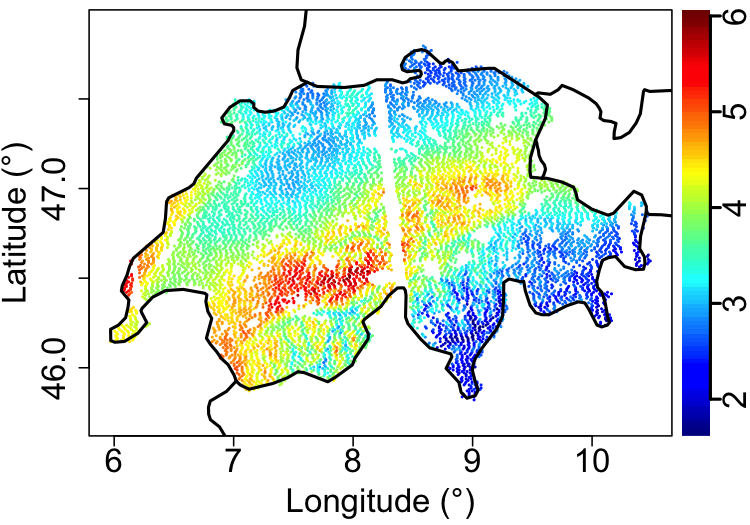}
    \caption{}
    \end{subfigure}
    \begin{subfigure}[c]{0.33\textwidth}
    \centering
    \includegraphics[keepaspectratio,height=3.5cm]{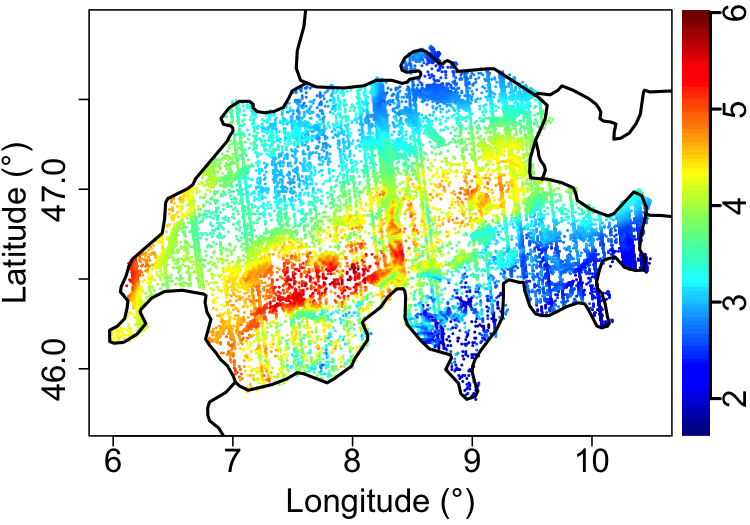}
    \caption{}
    \end{subfigure}
    \caption{Precipitation under the training Dense (a) and Sparse (b) training datasets, and test dataset (c).}
    \label{fig:illu_datasets}
\end{figure}

\begin{figure}[!h]
    \centering
    \begin{subfigure}[c]{0.33\textwidth}
    \centering
    \includegraphics[keepaspectratio,height=3.5cm]{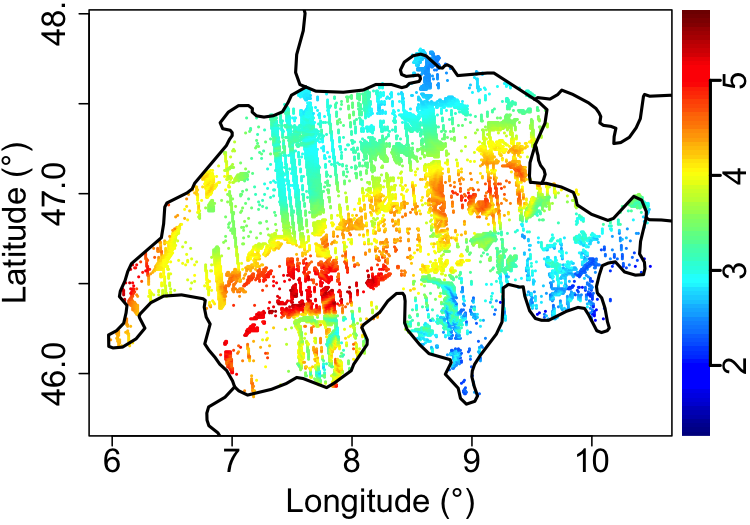}
    \caption{}
    \end{subfigure}
    \begin{subfigure}[c]{0.33\textwidth}
    \centering
    \includegraphics[keepaspectratio,height=3.5cm]{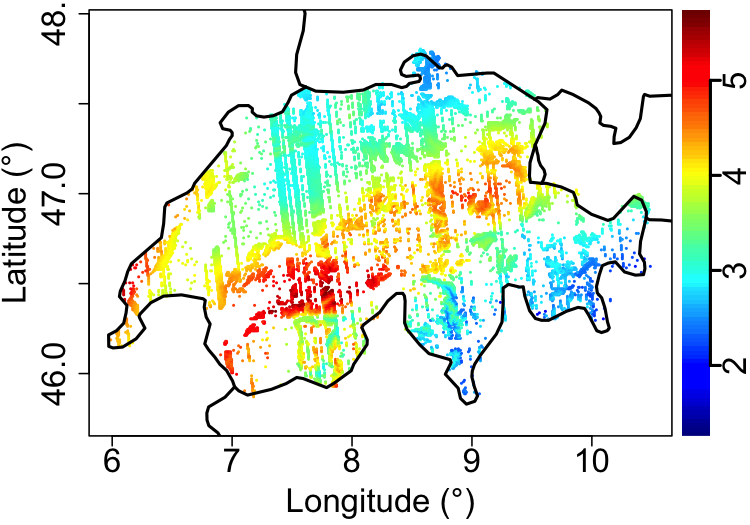}
    \caption{}
    \end{subfigure} \\
        \centering
    \begin{subfigure}[c]{0.33\textwidth}
    \centering
    \includegraphics[keepaspectratio,height=3.5cm]{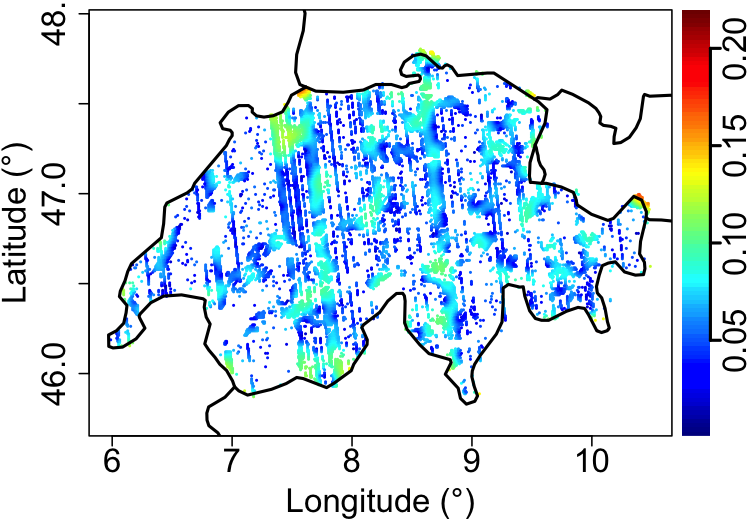}
    \caption{}
    \end{subfigure}
    \begin{subfigure}[c]{0.33\textwidth}
    \centering
    \includegraphics[keepaspectratio,height=3.5cm]{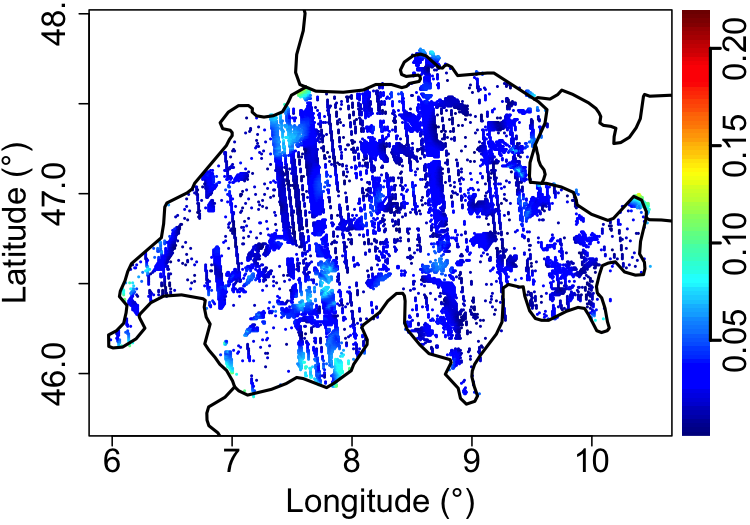}
    \caption{}
    \end{subfigure}
    \caption{Predictive means and standard deviations for M-STAT (first column), and M-NS (second column).}
    \label{fig:dense_preds}
\end{figure}

\begin{figure}[!h]
    \centering
    \begin{subfigure}[c]{0.33\textwidth}
    \centering
    \includegraphics[keepaspectratio,height=3.5cm]{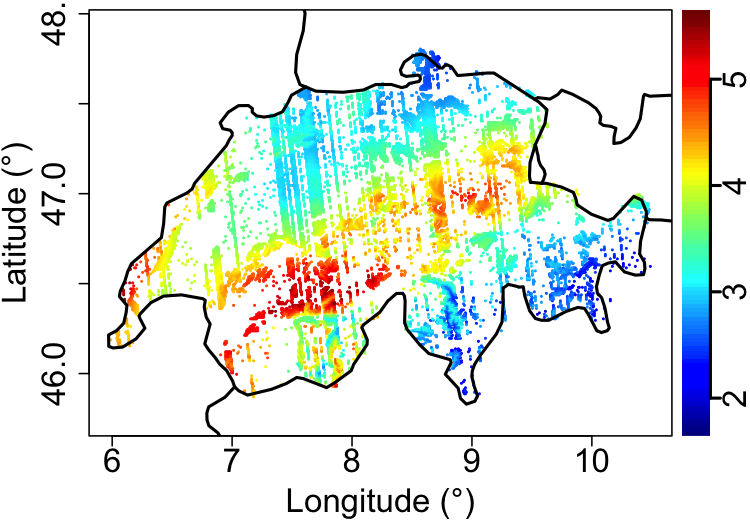}
    \caption{}
    \end{subfigure}
    \begin{subfigure}[c]{0.33\textwidth}
    \centering
    \includegraphics[keepaspectratio,height=3.5cm]{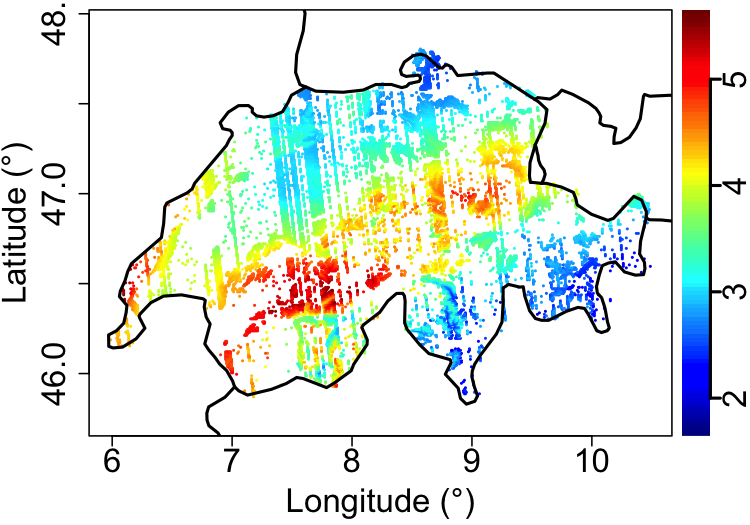}
    \caption{}
    \end{subfigure} \\
        \centering
    \begin{subfigure}[c]{0.33\textwidth}
    \centering
    \includegraphics[keepaspectratio,height=3.5cm]{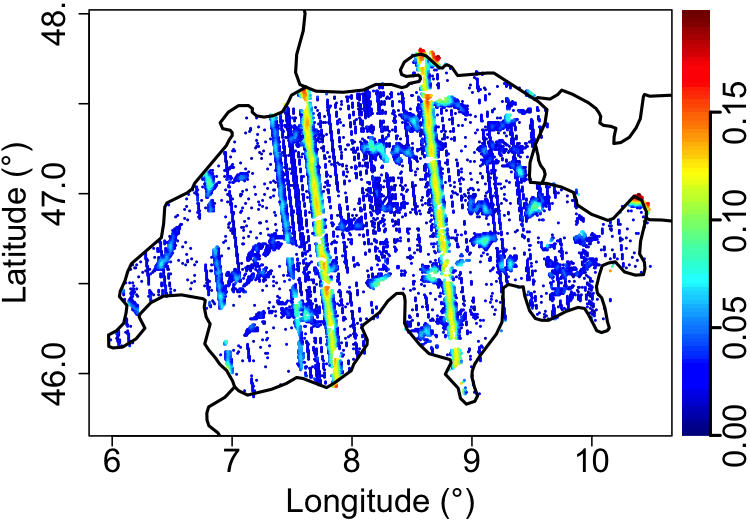}
    \caption{}
    \end{subfigure}
    \begin{subfigure}[c]{0.33\textwidth}
    \centering
    \includegraphics[keepaspectratio,height=3.5cm]{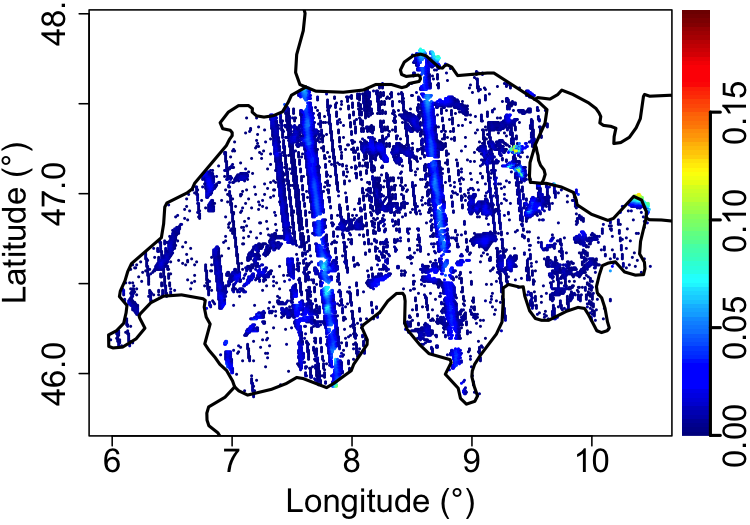}
    \caption{}
    \end{subfigure}
    \caption{Predictive means and standard deviations for M-STAT-T (first column), and M-NS-T (second column)}
    \label{fig:dense_preds}
\end{figure}

\subsection{Numerical summary of nonstationary models} \label{ap:model_tables}

The following tables present parameter estimates for M-NS and M-NS-T from the illustration section.

\begin{table}[h!]
    \centering
    \caption{Models hyperparameters.}
    \begin{tabular}{c|rrrrrrr}
         Model & $\lambda_r$ & $\lambda_{\beta}$ &  $\lambda_{\Sigma}$ & $\nu_{\text{min}}$& $\nu_{\text{max}}$ & $\delta$ & Taper function  \\ \hline
         M-STAT & $0.00$ & $0.00$ & $0.00$ & $0.5$ & $2.0$ & -- & -- \\
         M-NS & $0.01$ & $0.10$ & $0.20$ & 0.5 & 2.0 & -- & --  \\
         M-STAT-T & $.0125$ & $0.00$ & $0.00$ & $0.18$ & $1.5$ & $0.18$ & $\text{Wendland}_1$ \\
         M-NS-T & $0.01$ & $0.05$ & $0.40$ & $0.5$ & $2.5$ & $0.18$ & $\text{Wendland}_2$ \\
    \end{tabular}
    \label{tab:my_label}
\end{table}

\begin{table}[h!]
    \centering
    \caption{Parameter estimates for the dense model M-NS. Where available, the square root of the inverse of the Hesssian is given in parentheses. Slots where a (--) is present mean that the covariate was not considered in the final model.}
    \begin{tabular}{|c|rrrrrr}
    covariate & \verb|spat. mean| & \verb|std.dev| & \verb|scale| & \verb|aniso| & \verb|tilt| & \verb|smooth|\\ \hline
    \verb|intercept| & 3.396 \textit{(0.317)} & 0.934 \textit{(0.185)} & 0.876 \textit{(0.0091)} & -0.253 \textit{(0.047)} & 0.022 \textit{(0.062)} & 1.285 \textit{(0.199)} \\ 
    \verb|wind| & -0.408 \textit{(0.044)} & -- & -- & -- & -- & --  \\ 
    \verb|merwind| & 0.034 \textit{(0.031)} & -- & -- & -- & -- & --  \\ 
    \verb|BIO04| & -- & -- & -0.069 \textit{(0.017)} & -- & -- & --  \\ 
    \verb|BIO15| & -0.311 \textit{(0.035)} & -- & 0.044 \textit{(0.011)} & -- & -- & --  \\ 
    \verb|cloud| & -- & -- & -- & -- & -- & --  \\ 
    \verb|elev| & 0.025 \textit{(0.015)} & -- & -- & -- & -- & --  \\ 
    \verb|lati| & -- & -- & -- & -- & -- & -- \\ 
    \verb|long| & -- & -- & -- & -- & -- & --  \\         
    \verb|log(elev)| & -- & -0.086 \textit{(0.025)} & -- & 0.018 \textit{(0.011)} & 0.015 \textit{(0.016)} & --  \\ 
    \verb|log(cloud)| & -- & -- & -0.046 \textit{(0.015)} & -0.002 \textit{(0.029)} & -0.104 \textit{(0.032)} & --  \\ 
    \verb|log(wind)| & -- & -- & 0.064 \textit{(0.014)} & -- & -- & -- \\ 
    \verb|log(lati)| & -- & -- & -- & -- & -- & --  \\ 
    \verb|log(long)| & -- & -- & 0.095 \textit{(0.023)} & -- & -- & -- \\ 
    \end{tabular}
    \end{table}

    \begin{table}[h!]
        \centering
        \caption{Parameter estimates for the sparse model M-NS-T. Where available, the square root of the inverse of the Hesssian is given in parentheses. Slots where a (--) is present mean that the covariate was not considered in the final model.}
        \begin{tabular}{|c|rrrr}
            covariate & \verb|spat. mean| & \verb|std.dev| & \verb|scale| & \verb|smooth| \\
            \hline
            \verb|intercept| & 3.700 \textit{(0.007)} & -3.999 \textit{(0.016)} & 0.452 \textit{(0.039)} & 1.270 \textit{(0.043)}\\
            \verb|wind| & -0.384 \textit{(0.005)} & -- & -- & --  \\
            \verb|merwind| & 0.096 \textit{(0.007)} & -- & -- & --  \\
            \verb|BIO04| & --  & -- & -- & --  \\
            \verb|BIO15| & -0.440 \textit{(0.006)} & -0.215 \textit{(0.01)} & 0.114 \textit{(0.008)} & --  \\
            \verb|cloud| & -0.022 \textit{(0.007)} & -- & -- & -- \\
            \verb|elev| & 0.001 \textit{(0.001)} & -- & -- & --  \\
            \verb|lati| & -0.100 \textit{(0.008)} & -- & -- & --  \\
            \verb|long| & -0.328 \textit{(0.008)} & -- & -- & --  \\
            \verb|log(elev)| & -- & -- & -0.001 \textit{(0.002)} & --  \\
            \verb|log(cloud)| & -- & -- & -- & --  \\
            \verb|log(wind)| & -- & -0.157 \textit{(0.01)} & 0.402 \textit{(0.007)} & --  \\
            \verb|log(lati)| & -- & -- & -- & --  \\
            \verb|log(long)| & -- & -- & -- & --  \\
        \end{tabular}
    \end{table}
    
\subsection{Correlation functions with compact support}\label{sec:A.taper_covs}
{\tabcolsep0pt\begin{tabular}{lrl}
$\text{Spherical}$:\hspace*{5mm} & $\rho(h;\delta)$ & ${\displaystyle\; = I_{\{h<\delta\}}\Bigl(1- \frac{3}{2} \frac{h}{\delta} + \frac{1}{2}\frac{h^3}{\delta^3}\Bigr)}$, \\[3mm]
$\text{Wendland}_1$: & $\rho(h;\delta)$ & ${\displaystyle \;= I_{\{h<\delta\}}\Bigl(1- \frac{h}{\delta}\Bigr)^4 \Bigl(\frac{4h}{\delta} + 1 \Bigr)}$, \\[3mm]
$\text{Wendland}_2$: & $\rho(h;\delta)$ & ${\displaystyle \;= I_{\{h<\delta\}}\Bigl(1- \frac{h}{\delta}\Bigr)^6 \Bigl(\frac{35}{3} \frac{h^2}{\delta^2} + 6 \frac{h}{\delta} +1 \Bigr)}$,
\end{tabular}\\[3mm]}
where $\delta>0$.

\subsection{Details for accessing the dataset in Section~\ref{ch:illu}} \label{ap:access}

The dataset used in the Illustration is openly available via the web interface \url{https://doi.org/10.24381/cds.fe90a594}. 
It requires login credentials (non-academic credentials work as well).
Steps to download the intended dataset, as well as the preprocessing to select these covariates from Switzerland, and to retrieve elevation information, are available in the Git repository in \ref{sec:git}.

\subsection{Computational resources}
For the illustration and simulation studies, we used a Macbook Air M2 with 16Gb of memory RAM with macOS Sequoia 15.5.0.

\subsection{Source files}\label{sec:git}

R source files are available in the git repository \url{https://github.com/blasif/j.environ.2024}. 
The README.txt file gives an overview of the available files as well as how to run them. 

\subsection*{CRediT author statement}
\textbf{Federico Blasi}: Conceptualization, Methodology, Software, Formal Analysis, Data Curation, Writing - Original Draft, Writing - Review \& Editing, Visualization, \textbf{Reinhard Furrer}: Supervision, Writing - Review \& Editing. 

\section*{Acknowledgments}
The authors thank the reviewer and the editors for their for their very helpful comments and suggestions, which has greatly improved the overall quality of the article.
This work is supported by the Swiss National Science Foundation SNSF-175529. 

\bibliographystyle{plainnat}
\bibliography{references}

\end{document}